\documentclass[citeautoscript,aps,prb,twocolumn,superscriptaddress,longbibliography]{revtex4}

\usepackage{filecontents}
\usepackage[colorlinks=true,
linkcolor=blue,
urlcolor=blue,
citecolor=blue]{hyperref}

\usepackage{bm}
\usepackage[utf8]{inputenc}
\usepackage{physics}
\usepackage{natbib}
\usepackage{graphicx}
\usepackage{amsmath}
\usepackage[dvipsnames]{xcolor}
\usepackage{multirow}
\usepackage{soul}
\graphicspath{{images/}}

\begin{document}

\title{Rotational $g$ factors and Lorentz forces of molecules and solids from density-functional perturbation theory}

\author{Asier Zabalo} 
\affiliation{Institut de Ci\`encia de Materials de Barcelona (ICMAB-CSIC), Campus UAB, 08193 Bellaterra, Spain}
\author{Cyrus E. Dreyer}
\affiliation{Department of Physics and Astronomy,
             Stony Brook University,
             Stony Brook, New York, 11794-3800, USA}
\affiliation{Center for Computational Quantum Physics,
             Flatiron Institute,
             162 5th Avenue, New York, New York 10010, USA.}
\author{Massimiliano Stengel}
\affiliation{Institut de Ci\`encia de Materials de Barcelona (ICMAB-CSIC), Campus UAB, 08193 Bellaterra, Spain}
\affiliation{ICREA-Instituci\'o Catalana de Recerca i Estudis Avançats, 08010 Barcelona, Spain}

\date{\today}

\begin{abstract}
Applied magnetic fields can couple to atomic displacements via generalized Lorentz forces, which are commonly expressed as gyromagnetic $g$ factors. We develop an efficient first-principles methodology based on density-functional perturbation theory to calculate this effect in both molecules and solids to linear order in the applied field. Our methodology is based on two linear-response quantities: the macroscopic polarization response to an atomic displacement (i.e., Born effective charge tensor), 
and the antisymmetric part of its first real-space moment (the symmetric part corresponding to the dynamical quadrupole tensor). The latter quantity is calculated via an analytical expansion of the current induced by a long-wavelength phonon perturbation, and compared to numerical derivatives of finite-wavevector calculations. We validate our methodology in finite systems by computing the gyromagnetic $g$ factor of several simple molecules, demonstrating excellent agreement with experiment and previous density-functional theory and quantum chemistry calculations. In addition, we demonstrate the utility of our method in extended systems by computing the energy splitting of the low-frequency transverse-optical phonon mode of cubic SrTiO$_3$ in the presence of a magnetic field.
\end{abstract}

\maketitle

\section{Introduction}
An applied magnetic field has a significant impact on the lattice dynamics of
molecules and solids via generalized Lorentz forces, which are commonly
expressed as gyromagnetic $g$ factors \cite{PhysRevLett.89.116402}.
These are of great fundamental interest 
as manifestations of ``geometric magnetization'' \cite{PhysRevB.100.054408},
and enjoy an elegant formulation in terms of geometric phases \cite{PhysRevB.75.161101}
and Berry curvatures \cite{PhysRevB.86.104305}.
They are also related to the angular momentum of phonons via
the so-called ``phonon Zeeman effect'' \cite{PhysRevMaterials.3.064405,PhysRevMaterials.1.014401},
and are a crucial ingredient in the theory of the 
phonon Hall effect \cite{PhysRevB.86.104305,PhysRevB.103.214301,PhysRevLett.105.225901} (PHE).
In recent years significant advances have been made in the theoretical 
understanding of Lorentz forces in real systems,
\cite{PhysRevB.75.161101,culpitt2021ab,peters2021ab} but an
accurate and computationally efficient formalism for both molecules and extended crystals is 
still lacking. 

First-principles electronic-structure methods have traditionally 
been highly successful at calculating molecular $g$ factors. 
Reference values with 
chemical accuracy have been obtained long ago
in the context of
post–Hartree–Fock \emph{ab initio} methods, 
like coupled-cluster (CC) or M\/oller–Plesset (MP) 
perturbation theory \cite{cybulski1997calculation}.
The works by Ceresoli and Tosatti \cite{PhysRevLett.89.116402,ceresoli2002berry}
later demonstrated that density functional theory (DFT) can provide reliable values
at a significantly lower computational cost; also, their pioneering Berry-phase approach 
has paved the way towards the development of the 
``modern theory of magnetization'' \cite{PhysRevB.74.024408,PhysRevLett.99.197202,resta2010electrical}.

The case of extended solids has been comparatively much less explored.
The reason is that previous approaches required performing calculation in the presence of a finite external magnetic field (${\bf B}$), which is a challenge to incorporate with periodic boundary conditions. Though there has been theoretical work in this direction \cite{PhysRevLett.92.186402,lee2007electronic}, so far a widespread implementation is lacking.
This situation is in stark contrast with the case of an isolated molecules,
where finite-${\bf B}$ methods are well established in existing codes \cite{Li2019_e1436,balasubramani2020turbomole}. 
As a result, reference theoretical values for the coupling constants between
phonons in solids and an external magnetic field are still scarce.
Recent works by Spaldin and coworkers \cite{PhysRevMaterials.3.064405,PhysRevMaterials.1.014401}, 
do report 
first-principles values for the phonon $g$ factors in 
a broad range of crystalline insulators; 
however, a point-charge model for the microscopic currents associated 
with the ionic orbits was assumed therein.
This certainly constitutes a drastic simplification from the computational perspective,
as it only requires calculating standard linear-response properties (e.g., the Born
effective charge tensor); however, the validity of such an approximation has not been tested yet. 

Here we establish, in the framework of first-principles density-functional perturbation theory (DFPT),
an accurate and computationally efficient methodology to compute both generalized Lorentz forces 
and $g$ factors in molecules \emph{and} solids.
Our strategy consists in defining both quantities
in terms of the microscopic electronic and nuclear currents, ${\bf J}({\bf r})$, that accompany 
the adiabatic evolution of the system along the atomic trajectories.
In particular, the first spatial moment of ${\bf J}({\bf r})$ can be regarded as
a geometric orbital magnetic moment, ${\bf m}$, 
which couples linearly to the 
external ${\bf B}$ field and acts as an effective vector potential in the 
classical ionic Lagrangian.
At the leading order in the ionic velocities, ${\bf v}$, the calculation of ${\bf m}$
can be carried out in the framework of density-functional perturbation theory
via a long-wave expansion of the macroscopic polarization response to 
a phonon. 
Such expansion, in turn, is written in terms of two
linear-response tensors:~\cite{Stengel2013a} the macroscopic polarization induced by an atomic 
displacement $\mathbf{J}^{(0)}$, corresponding to the Born effective charges (BECs), 
and its first-order spatial dispersion, $\mathbf{J}^{(1)}$. 
BECs are routinely calculated in many publicly-available density-functional theory 
codes \cite{PhysRevB.55.10355,PhysRevB.43.7231,PhysRevB.58.6224}; 
the main technical challenge resides then in the calculation of $\mathbf{J}^{(1)}$.

In the course of this work we have implemented and used two different approaches 
for accessing $\mathbf{J}^{(1)}$, and compared their mutual consistency as part of our numerical 
tests.
The first method, based on Ref. ~\onlinecite{Dreyer2018}, consists in performing the DFPT 
calculations of the polarization response at finite ${\bf q}$, and subsequently taking their long-wave 
expansion via numerical differentiation. 
The second method, which we shall prefer from the point of view of 
computational convenience, consists in taking the long-wave expansions
analytically via the recently implemented~\cite{Royo_Stengel_PRX,romero-20} 
long-wave module of {\sc abinit}.~\cite{gonze2009abinit,gonze2020abinit}
Note, however, that the existing implementation only works for the 
\emph{symmetric} part of $\mathbf{J}^{(1)}$, corresponding to 
the dynamical quadrupole tensor, while for the present purposes 
we require the \emph{antisymmetric} part of the tensor,
which has not been addressed earlier. 
For its implementation, we have further extended
the capabilities of {\sc abinit} by incorporating the wavefunction response to an
orbital ${\bf B}$ field. One can show that the resulting formulation of the geometric orbital magnetization
 nicely recovers the theory of Ref. ~\onlinecite{PhysRevB.100.054408}, including the additional \textit{topological} contribution derived therein.

To demonstrate our method, we first consider the gyromagnetic $g$ factor,
which depends on the magnetic moment that is associated with a uniform and rigid 
rotation of a finite body. 
We show that our formula, based on the calculation of $\mathbf{J}^{(1)}$, consistently
yields a vanishing magnetic moment in the case of a neutral closed-shell 
atom, and correctly transforms upon a change of the assumed center of rotation.
Our numerical results for several representative molecules show excellent agreement with experiment 
and with earlier calculations, where available; the elements of $\mathbf{J}^{(1)}$ that we obtained 
via either finite-difference or analytical long-wave expansions nicely match in all tested cases.
For comparison, we also test an alternative formulation, based on a coordinate transformation
to the co-moving frame of the rotating molecule,~\cite{PhysRevB.98.125133} and discuss its 
performance regarding numerical convergence and other technical issues (e.g., related to the use of 
nonlocal pseudopotentials).

Next, we consider the magnetization induced by a circularly polarized optical phonon, which 
we express as a generalized Lorentz force in presence of a uniform magnetic field.
As a physical manifestation of this effect, we calculate the splitting of the soft polar transverse-optical (TO)
mode frequencies of SrTiO$_3$ at the Brillouin zone center due to an external magnetic field.
Our motivation for revisiting this system comes the very recent measurement of a giant phonon Hall effect~\cite{PhysRevLett.124.105901} in the same material.
As in the case of the molecular $g$ factors,
we base our discussion on the calculation of
the $\mathbf{J}^{(1)}$ tensor, which we perform both via the approach of Ref. ~\onlinecite{Dreyer2018},
and via the analytical long-wave method; again, we find excellent numerical agreement between
the two.

The remainder of the paper is organized as follows.  Sec.~\ref{Sec_Theory} and 
\ref{Sec_implementation} are devoted to introducing the formalism and computational implementation for calculating molecular $g$ factors and 
generalized Lorentz forces in extended solids.
In Sec.~\ref{Sec_Results} we present results on the gyromagnetic $g$ factors of some simple molecules and the computation of the 
generalized Lorentz force in cubic SrTiO$_3$. The latter enables the calculation of the frequency splitting of the TO modes in presence of a magnetic field. We conclude the paper with Sec.~\ref{Sec_conclusions}.

\section{Theory}\label{Sec_Theory}
\subsection{Lagrangian for a solid under an applied magnetic field}
Consider the nonadiabatic Ehrenfest Lagrangian of the crystalline system under an 
applied magnetic field
\begin{equation}
\label{ehrenfest}
\begin{split}
\mathcal{L}=&\sum_{l\kappa\alpha} \frac{1}{2}M_\kappa(\dot{R}^l_{\kappa\alpha})^2 
+\sum_{l\kappa}Z_\kappa \dot{\mathbf{R}}_\kappa^l\cdot\mathbf{A}(\mathbf{R}^l_\kappa) \\
&+\sum_j \bra*{\phi_j}[i\partial_t -H_\text{el}(\mathbf{A},\{\mathbf{R}^l_{\kappa}\})
]\ket*{\phi_j}
\end{split}
\end{equation}
where $\mathbf{A}$ is the magnetic vector potential, 
${\bf R}^l_{\kappa}$ represents the position of ion $l\kappa$ within the crystal ($\kappa$ is a basis index and $l$ refers to the cell),  
$M_\kappa$ is the mass of ion $\kappa$ and $Z_\kappa$ its bare (pseudo-)charge.
Regarding the electronic part, $\phi_j$ are the Kohn-Sham orbitals and  $H_\text{el}$ is the electronic Hamiltonian, 
depending parametrically on the ionic positions,	
\begin{equation}\label{Eq_H_el}
H_\text{el}(\mathbf{A},\{\mathbf{R}^l_\kappa \})= \frac{1}{2} [ \mathbf{p}+\mathbf{A}(\mathbf{r})]^2+
V_\text{eff}(\{\mathbf{R}^l_\kappa\}).
\end{equation} 
(We use Hartree atomic units, i.e., the electron mass and charge are $m_e=1$ and $-e=-1$, respectively.)
If we assume that the external magnetic fields are small (an excellent approximation in the
vast majority of cases), we can work at linear order in the vector potential and write
\begin{equation}
\begin{split}
\mathcal{L}=&\sum_{l\kappa\alpha} \frac{1}{2}M_\kappa(\dot{R}^l_{\kappa\alpha})^2 
+\sum_{l\kappa}Z_\kappa \dot{\mathbf{R}}_\kappa^l\cdot\mathbf{A}(\mathbf{R}^l_\kappa) \\
&+\sum_j \bra*{\phi_j}[i\partial_t -H_\text{el}(\{\mathbf{R}^l_{\kappa}\}) ]\ket*{\phi_j} \\
&+ \int d^3 r {\bf A}({\bf r})\cdot {\bf J}^{\rm el}({\bf r}),
\end{split}
\end{equation}
where the microscopic electronic currents (in zero external field) are defined as
\begin{equation}
{\bf J}^{\rm el}({\bf r}) = -\sum_j \frac{1}{2} \bra*{\phi_j} \Big( {\bf p}|{\bf r}\rangle \langle {\bf r} | + 
|{\bf r}\rangle \langle {\bf r} |  {\bf p} \Big)  \ket*{\phi_j}.
\end{equation}
As we treat the nuclei as classical point charges, the ionic currents read as
\begin{equation}
{\bf J}^{\rm ion}({\bf r}) = \sum_{l\kappa} Z_\kappa \dot{\mathbf{R}}_\kappa^l \delta({\bf r} - \mathbf{R}_\kappa^l);
\end{equation}
this allows us to reabsorb the effects of the external vector potential in
a single interaction term,
\begin{equation}
\begin{split}
\mathcal{L}=&\sum_{l\kappa\alpha} \frac{1}{2}M_\kappa(\dot{R}^l_{\kappa\alpha})^2  \\
&+\sum_j \bra*{\phi_j}[i\partial_t -H_\text{el}(\{\mathbf{R}^l_{\kappa}\}) ]\ket*{\phi_j} \\
& + \int d^3 r {\bf A}({\bf r})\cdot {\bf J}({\bf r}),
\end{split}
\end{equation}
where ${\bf J}= {\bf J}^{\rm el} + {\bf J}^{\rm ion}$.
By choosing the symmetric gauge, $\mathbf{A}=\frac{1}{2}\mathbf{B}\times \mathbf{r}$, 
we can equivalently write
\begin{equation}
\int d^3 r {\bf A}({\bf r})\cdot {\bf J}({\bf r}) = {\bf B}\cdot {\bf m}(\{\mathbf{R}^l_{\kappa}\}, \{\dot{\mathbf{R}}^l_{\kappa}\} ),
\end{equation}
where ${\bf m} = \int d^3 r {\bf r} \times {\bf J} /2$ is the geometric magnetic moment
associated with the dynamical evolution of the ions along their trajectories.
We are now ready to take the adiabatic approximation, in a regime where the
ionic velocities are small,
\begin{equation}\label{Eq_L_system}
\begin{split}
\mathcal{L}=&\sum_{l\kappa\alpha}\frac{1}{2}M_\kappa(\dot{R}^l_{\kappa\alpha})^2 
 +  {\bf B}\cdot \sum_{l\kappa\alpha} \dot{R}^l_{\kappa\alpha} {\bf m}^l_{\kappa \alpha} (\{\mathbf{R}^l_{\kappa}\}) \\
 & -E_\text{KS}(\{\mathbf{R}^l_{\kappa}\}),
\end{split}
\end{equation}
where the two new terms are the Born-Oppenheimer potential energy surface
in zero field, $E_\text{KS}$, plus a term that depends on the \emph{dynamical orbital 
magnetic moment tensor},
\begin{equation}\label{m_l_kalpha}
\mathbf{m}^l_{\kappa \alpha} = \frac{\partial {\bf m}}{\partial \dot{R}^l_{\kappa\alpha}} \Big|_{\dot{R}^l_{\kappa\alpha}=0}.
\end{equation}
The latter quantity differs to the Born effective charge (BEC) tensor in that
the adiabatic macroscopic ${\bf m}$, rather than the adiabatic macroscopic
current ${\bf J}$, is differentiated with respect to the ionic velocities.
Note that ${\bf m}^l_{\kappa \alpha}$ generally depends on the electromagnetic gauge,
unlike the BEC; however, as we shall see shortly, its consequences on 
ionic dynamics are gauge-independent.
This is a common feature of physical problems that involve an applied external ${\bf B}$;
and indeed, the velocity-dependent potential
\begin{equation}
\label{Eff_vec_pot}
\tilde{A}^{l}_{\kappa \alpha} = {\bf B}\cdot{\bf m}^l_{\kappa \alpha}
\end{equation} 
can be regarded as an effective  
vector potential, $\tilde{\bf A}^{l}_{\kappa}$, acting on the ion $l\kappa$,
and whose magnitude depends on the specific atom under consideration.  
This leads to the following expression for the classical Hamiltonian
of the ions,
\begin{equation}\label{Eq_H_system}
\begin{split}
\mathcal{H}=&\sum_{l\kappa\alpha}\frac{1}{2}M_\kappa[\dot{R}^l_{\kappa\alpha} - \tilde{A}^{l}_{\kappa \alpha}(\{\mathbf{R}^l_{\kappa}\})]^2 
+ E_\text{KS}(\{\mathbf{R}^l_{\kappa}\}),
\end{split}
\end{equation}
which is good up to linear order in the ionic velocities,
and where the vector potential emerges from the breakdown of time-reversal
symmetry (TRS) that is associated with the external ${\bf B}$.
One can show that this treatment is fully consistent with 
the conventional expression, \cite{cc082398dfa9424b87ec93ba7b5c4305} 
where $\tilde{A}^{l}_{\kappa \alpha}$ is written as a
Berry connection in the parameter space of the ionic coordinates.
The advantage of the present formulation rests on the availability of
efficient first-principles methods to compute directly ${\bf m}^l_{\kappa \alpha}$,
and hence the vector potential $\tilde{A}^{l}_{\kappa \alpha}$, without
the need of incorporating an external ${\bf B}$ field in the simulation.
We shall discuss this point in the next subsection.

\subsection{Geometric magnetization}\label{Sec_preliminaries}

The basic quantity we shall be dealing with is the microscopic polarization
response to the displacement of an isolated atom,~\cite{Stengel2013a}
\begin{equation}\label{Eq_P}
\bm{\mathcal{P}}_{\kappa\beta}(\mathbf{r}- {\bf R}^l_{\kappa})=\frac{ \partial \mathbf{J} ({\bf r}
  )}{\partial\dot{R}^l_{\kappa\beta}}.
\end{equation}
Eq.~(\ref{Eq_P}) always sets the coordinate origin to the atomic site; therefore, 
the functions $\bm{\mathcal{P}}_{\kappa\beta}(\mathbf{r})$ do not depend on the cell 
index $l$.
[Recall that $l$ runs over all the unit cells, and $\mathbf{R}^l_\kappa=\mathbf{R}^l+\boldsymbol{\tau}_\kappa$, where
$\mathbf{R}^l$ is a Bravais lattice vector and $\boldsymbol{\tau}_\kappa$ is the position 
of ion $\kappa$ within the unit cell.]
Note that the vector fields
contain both electronic and ionic contributions, i.e.,
\begin{equation}
\mathcal{P}_{\alpha,\kappa\beta}(\mathbf{r})=\mathcal{P}^{\text{el}}_{\alpha,\kappa\beta}(\mathbf{r})+
\mathcal{P}^{\text{ion}}_{\alpha,\kappa\beta}(\mathbf{r}),
\end{equation} 
where the $\alpha$ subscript indicates the Cartesian component. 
The ionic contribution comes in the form of a Dirac delta function that carries 
the bare nuclear (or pseudopotential) charge $Z_\kappa$,
\begin{equation}
\mathcal{P}^{\text{ion}}_{\alpha,\kappa\beta}(\mathbf{r})=Z_\kappa \delta_{\alpha\beta}
\delta(\mathbf{r}). 
\end{equation}

For most practical purposes, it is convenient to  
expand the microscopic polarization field into a multipole series, by writing the lowest-order moments as 
\begin{equation}\label{Eq_P_series}
\begin{split}
J^{(0)}_{\alpha,\kappa\beta}&=\int d^3r \, \mathcal{P}_{\alpha,\kappa\beta}(\mathbf{r}),\\
J^{(1,\gamma)}_{\alpha,\kappa\beta}&=
\int d^3r \, r_\gamma \mathcal{P}_{\alpha,\kappa\beta}(\mathbf{r}).
\end{split}
\end{equation}
(Note that, in order to ensure the convergence of the above 
integrals, some care is required in the treatment of the macroscopic
electric fields; techniques to deal with this issue are now 
well established \cite{Martin1972,Stengel2013a}.)
$\mathbf{J}^{(0)}$ corresponds to the Born effective charge tensor and 
$\mathbf{J}^{(1)}$ is the first moment of the polarization response, whose symmetric part
corresponds to the dynamical quadrupole tensor \cite{Stengel2013a,Royo_Stengel_PRX}
\begin{equation}\label{Eq_Quadrupole}
Q^{(2,\alpha\gamma)}_{\kappa\beta}=J^{(1,\gamma)}_{\alpha,\kappa\beta}+
J^{(1,\alpha)}_{\gamma,\kappa\beta}.
\end{equation}
On the other hand, the antisymmetric part of $\mathbf{J}^{(1)}$ contributes to the
magnetization response to the atomic velocity, and can be expressed as
\begin{equation}
  \label{Eq_bigM}
\mathcal{M}_{\alpha,\kappa\beta}=\frac{1}{2}\sum_{\gamma,\delta}\epsilon^{\alpha\gamma\delta}
J^{(1,\gamma)}_{\delta,\kappa\beta},
\end{equation}
where $\epsilon^{\alpha\gamma\delta}$ is the Levi-Civita symbol. 
More precisely, $\bm{\mathcal{M}}$ is the magnetic
moment of the electronic currents calculated with respect to the
unperturbed atomic position, which follows from the definition of
$\mathbf{J}^{(1)}$ in Eq. (\ref{Eq_P_series}).

The above definitions lead to the following formula for the geometric 
magnetic moment associated with the adiabatic motion of the ion $l\kappa$,
\begin{equation}
\label{mlk}
\begin{split}
{\bf m}^l_{\kappa \beta} &=  \frac{1}{2} \int d^3 r {\bf r} \times \bm{\mathcal{P}}_{\kappa\beta}(\mathbf{r}-{\bf R}_\kappa^l) \\
 &= \frac{1}{2} {\bf R}^l_{\kappa} \times {\bf Z}_{\kappa \beta} + \bm{\mathcal{M}}_{\kappa\beta},
\end{split}
\end{equation}
where $Z^*_{\alpha,\kappa \beta} = J^{(0)}_{\alpha,\kappa\beta}$ is the $\alpha$ component
of the polarization induced by a displacement of atom $\kappa$ along $\beta$, i.e.,
the Born effective charge.
This expression clarifies the gauge-dependence of ${\bf m}^l_{\kappa \beta}$ that
we have anticipated in the previous subsection: this quantity depends explicitly 
on the absolute atomic position, and hence on the arbitrary choice of the 
coordinate origin.

In the case of an isolated and neutral molecule, it is insightful to consider the sublattice sum
of ${\bf m}_{\kappa \beta}={\bf m}^0_{\kappa \beta}$, which corresponds physically to the
magnetic moment associated with a rigid translation of the body. Because of the acoustic sum rule, 
the origin indeterminacy disappears; then, by using the \emph{dipolar sum rule} of Appendix~\ref{app:dipolar},
we arrive at
\begin{equation}
\label{summk}
\sum_\kappa m_{\alpha,\kappa \beta} = \frac{1}{2}\sum_{\gamma} \epsilon^{\beta \alpha \gamma} \mathcal{D}_\gamma,
\end{equation}
where $\bm{\mathcal{D}}$ is the static dipolar moment of the molecule,
\begin{equation}
\bm{\mathcal{D}} = \int d^3 r \, {\bf r} \, \rho^{(0)}({\bf r}).
\end{equation}
Eq.(\ref{summk}) is precisely the expected result for the uniform rigid 
motion of a distribution of classical charges whose local
density equals $\rho^{(0)}({\bf r})$.

\subsection{Magnetization by rotation: rotational $g$ factors}\label{Sec_mag_rotation}
Consider an isolated molecule molecule to which we apply a time-dependent 
counter-clockwise rotation along the axis $b$ 
by an angle $\theta_b$.
In general, the magnetic moment
can be expressed as \cite{doi:10.1080/00268970009483366}
\begin{equation}
m_a =\frac{1}{2}\sum_j g_{aj}L_j,
\end{equation}
where $g_{aj}$ is the $g$ tensor. $L_j$ is the angular momentum, given by
\begin{equation}
L_j=\sum_b I_{jb}\omega_b,
\end{equation}
where $\mathbf{I}$ is the moment of inertia matrix.
($\omega_b= \dot{\theta}_b$ is the angular velocity, defined as
time derivative of the rotation angle.)
Thus,
\begin{equation}\label{ma_molecule_II}
\frac{\partial m_a}{\partial\omega_b}=\frac{1}{2}\sum_{j}g_{aj}I_{jb},
\end{equation}
In the reference frame where $\mathbf{I}$ is diagonal, the $g$ tensor can then be written as
\begin{equation}
  \label{Eq_gfac}
g_{ab}=\frac{2}{I_{bb}} \frac{\partial m_a}{\partial\omega_b}.
\end{equation}
We shall now derive a closed formula for
the  magnetic moment induced by a uniform rotation of the molecule.
We shall present two alternative results, the first calculated in the 
standard Cartesian frame based on the quantities introduced in 
the previous Section, and the second based on the comoving frame 
theory of Ref. ~\onlinecite{PhysRevB.98.125133}.

\subsubsection{Cartesian frame}

A rigid rotation about an arbitrary axis can be represented as 
the following displacement of the individual atoms,
\begin{equation}
\label{ukth}
\mathbf{u}_\kappa=\boldsymbol{\theta}
\cross\boldsymbol{\tau}_\kappa,
\end{equation}
where we have introduced the rotation pseudovector
$\boldsymbol{\theta}=\theta\hat{r}_b$.  
By combining Eq.~(\ref{ukth}) with Eq.~({\ref{mlk}}), the magnetic
moment associated with the rigid rotation of the sample can be expressed
in terms
of the dynamical magnetization and Born effective 
charge tensors defined in the previous subsection, 
\begin{equation}\label{Eq_mz}
\frac{\partial m_a}{\partial\omega_b} 
= \sum_{\kappa,j,\beta}\epsilon^{b j\beta}\tau_{\kappa j} 
\left[ \mathcal{M}_{a,\kappa\beta} +\frac{1}{2} 
 \sum_{i,\alpha}
\epsilon^{ai\alpha} \tau_{\kappa i} J^{(0)}_{\alpha,\kappa\beta} \right].
\end{equation}
This formula, containing the first moment of the dynamical magnetic
dipoles and the second moment of the dynamical electrical dipoles,
is valid only if the electromagnetic gauge origin coincides with
the center of rotation of the molecule; this ensures, via rotational
symmetry, that the linear-response result coincides with the average 
geometric magnetization accumulated in a cyclic loop. \cite{PhysRevB.98.125133}
In Appendix~\ref{app:shift} we shall prove that, upon a simultaneous shift 
of the gauge origin and center of rotation by ${\bf R}$, the above formula 
transforms as 
\begin{equation}
\label{gauge}
\frac{\partial m_a({\bf R})}{\partial\omega_b} - \frac{\partial m_a({\bf 0})}{\partial\omega_b} =
   \frac{R_a \mathcal{D}_b + R_b \mathcal{D}_a}{2} -   \delta_{ab} {\bf R} \cdot \bm{\mathcal{D}}.
\end{equation}
Thus, ${\partial m_a({\bf R})}/{\partial\omega_b}$ is origin-independent 
in nonpolar molecules (i.e., molecules with vanishing static dipole).
In other cases, the result depends on the assumed center of
rotation, which is usually set as the center of mass of the
system.

\subsubsection{Comoving frame}
 
By using the 
theory of Ref. ~\onlinecite{PhysRevB.98.125133}, the rotational geometric magnetization 
can be expressed as \cite{cybulski1994calculations}
\begin{equation}
\begin{split}\label{Eq_mz2}
\frac{\partial m_a}{\partial\omega_b} =& -2\chi^{\rm mag}_{ab} + \frac{1}{2}  \int d^3 r 
\frac{\partial [{\bf r} \times (\omega \times {\bf r})]_a}{\partial \omega_b} \rho^{(0)}({\bf r})\\
 =& -2\chi^{\rm mag}_{ab} + \frac{1}{2} \sum_{ij\beta} \epsilon^{ai\beta} \epsilon^{bj\beta} \int d^3 r r_i r_j \rho^{(0)}({\bf r})\\
 =& -2\chi^{\rm mag}_{ab} + \frac{1}{2} \int d^3 r (\delta_{ab} r^2 - r_a r_b) \rho^{(0)}({\bf r}).
\end{split}
\end{equation}
The first term is proportional to the magnetic susceptibility, and originates from the electronic currents
in the reference frame that is rigidly rotating with the sample; the second term describes the
magnetic moment generated by the rigid rotation of the ground-state charge density of
the molecule, and serves to convert the result
to the laboratory frame.
Upon a shift of the gauge origin, $\chi^{\rm mag}_{ab}$ remains unaltered 
while the second term trivially transforms as in Eq.~(\ref{gauge}).
(Clearly, the quadrupole becomes origin-dependent whenever a nonzero
dipolar moment is also present, consistent with the above arguments.)

As part of the validation of our implementation, we shall compute the
geometric magnetization by using both methods, Eq.~(\ref{Eq_mz}) and Eq.~(\ref{Eq_mz2}).
We can anticipate, however, that Eq.~(\ref{Eq_mz}) is preferable in practical
applications, for the following reasons.
First, the widespread use of nonlocal pseudopotentials is a concern in regards to 
Eq.~(\ref{translational}), which is a prerequisite for Eq.~(\ref{Eq_mz2}) to be
valid.
[In particular, the equivalence between Eq.~(\ref{Eq_mz}) and Eq.~(\ref{Eq_mz2})
rests on the translational invariance at the quadrupolar order, see the discussion 
around Eq.~(\ref{tra2}).]
Because of this issue, we find that Eq.~(\ref{Eq_mz2}) yields qualitatively
incorrect results for systems where ${\partial m_a}/{\partial\omega_b}$ must
vanish identically, e.g., in isolated noble gas atoms or molecular dimers that
rotate about their axis.
Second, even in cases where Eq.~(\ref{Eq_mz2}) is exact (e.g., in the H$_2$ molecule
whenever hydrogen is described by a local pseudopotential), its numerical implementation
involves the calculation of the static quadrupolar moment of the molecule, which might
converge slowly as a function of the cell size. (We shall illustrate this point 
in practice in Sec. \ref{Sec_Res_g}.)

\subsection{Magnetization induced by a circularly polarized optical phonon: generalized Lorentz force
  \label{Sec_circ_phonon}}
We now consider the case of a circularly polarized optical phonon describing a cyclic path along orbits in a given plane. 
In presence of time-reversal symmetry (TRS), the clockwise and counterclockwise orbits are
degenerate. Here, we take the approach of breaking 
TRS via an external $\mathbf{B}$ field
oriented along $\gamma$, and discuss the implications on lattice
dynamics within the harmonic regime of small displacements.

In order to compute the derivatives of the Lagrangian with respect to the ionic displacements (${\bf u}^l_\kappa$) and velocities 
($\dot{\bf u}^l_\kappa$), we expand the total orbital magnetic moment of the system up to first order in 
both ${\bf u}^l_\kappa$ and $\dot{\bf u}^l_\kappa$, and the Kohn-Sham energy up to second order 
in ${\bf u}^l_\kappa$ (harmonic approximation). 
The Lagrangian of Eq. (\ref{Eq_L_system}) then reads as
\begin{equation}
\begin{split}
\mathcal{L}&=\sum_{l\kappa\alpha}\frac{1}{2}M_\kappa(\dot{u}^l_{\kappa\alpha})^2 
+ \sum_{l\kappa\alpha } \Big(
\frac{\partial m_\gamma}{\partial \dot{u}_{\kappa\alpha}^l} \dot{u}^l_{\kappa\alpha}B_\gamma -
\frac{\partial E_\text{KS}}{\partial u^l_{\kappa\alpha}}u^l_{\kappa\alpha}\Big)
\\
&+ 
\sum_{\substack{l\kappa\alpha \\ l'\kappa'\beta}}\Big(
\frac{\partial^2 m_\gamma}{\partial u_{\kappa\alpha}^l\partial\dot{u}^{l'}_{\kappa'\beta}}u^l_{\kappa\alpha}\dot{u}^{l'}_{\kappa'\beta}B_\gamma -
\frac{\partial^2 E_\text{KS}}{\partial u^l_{\kappa\alpha}\partial u^{l'}_{\kappa'\beta}}u^l_{\kappa\alpha}u^{l'}_{\kappa'\beta}\Big).
\end{split}
\end{equation}
The first line consists, next to the kinetic term, in 
a constant vector potential field acting on individual ions, which can be 
gauged out; and in the static forces in the initial configuration, which 
we assume to vanish.
Based on these observations, we can now obtain the Euler-Lagrange equations of 
motion via 
\begin{equation}\label{Eq_Euler_Lagrange}
\frac{d}{dt}\frac{\partial\mathcal{L}}{\partial\dot{u}^0_{\kappa\alpha}}-
\frac{\partial\mathcal{L}}{\partial u^0_{\kappa\alpha}}=0,
\end{equation}
which leads to
\begin{equation}\label{Eq_motion_Phi}
\begin{split}
M_\kappa \ddot{u}^0_{\kappa\alpha}
=& -\sum_{l\kappa\beta}\Big[ \Phi_{\kappa\alpha,\kappa'\beta}(0,l)u^{l}_{\kappa'\beta}-
\Phi^\gamma_{\kappa\alpha,\kappa'\beta}(0,l) \dot{u}^{l}_{\kappa'\beta}B_\gamma  \Big],
\end{split}
\end{equation} 
Here $\mathbf{\Phi}$ is the usual real-space interatomic force-constant matrix and we have defined 
\begin{equation}\label{Eq_Phi_gamma_1}
\Phi^\gamma_{\kappa\alpha,\kappa'\beta}=
\frac{\partial^2 m_\gamma}{\partial u_{\kappa\alpha}\partial\dot{u}_{\kappa'\beta}}-\frac{\partial^2 m_\gamma}{\partial\dot{u}_{\kappa\alpha}\partial u_{\kappa'\beta}},
\end{equation}
which is the (antisymmetric) generalized Lorentz force produced by the external magnetic field. 
By using this result in combination with Eq.~(\ref{m_l_kalpha}) and Eq.~(\ref{mlk}), we obtain
\begin{equation} \label{Eq_full_phi}
\begin{split}
 \Phi^\gamma_{\kappa\alpha,\kappa'\beta} &= 
	 \Phi^{\rm pc,\gamma}_{\kappa\alpha,\kappa'\beta} +
     \Phi^{\rm di,\gamma}_{\kappa\alpha,\kappa'\beta} +
     \Phi^{\rm ea,\gamma}_{\kappa\alpha,\kappa'\beta}.
\end{split}
\end{equation}
The meaning of the three terms on the rhs goes as follows.
First, we have an on-site contribution 
that only depends on the Born dynamical charges,
\begin{equation} \label{pc}
\Phi^{\rm pc,\gamma}_{\kappa\alpha,\kappa'\beta} = \frac{1}{2}\delta_{\kappa\kappa'}\sum_{l}\Big(
\epsilon^{\gamma\alpha l}Z^*_{l,\kappa\beta}
-\epsilon^{\gamma\beta l}Z^*_{l,\kappa\alpha}
\Big).
\end{equation}
The ``point-charge'' (pc) denomination indicates that, in absence of 
electrons, the ${\bf Z}$ tensor becomes a constant, $Z^*_{l,\kappa \alpha} = Z_\kappa \delta_{\alpha l}$,
and Eq.~(\ref{pc}) reduces then 
to the well-known Lorentz 
force (L) acting on a classical test particle of charge $Z_\kappa$, 
\begin{equation}\label{Eq_Lorentz_particle}
\Phi^{\rm L,\gamma}_{\kappa\alpha,\kappa'\beta} = \delta_{\kappa \kappa'} Z_\kappa \epsilon^{\gamma \alpha \beta}.
\end{equation}
This term was described in Refs. ~\onlinecite{PhysRevMaterials.3.064405} and  ~\onlinecite{PhysRevMaterials.1.014401}.
Next, we have a ``dispersion'' (di) contribution, which stems from the 
fact that the electronic currents associated with ionic motion are 
spread out in space around the nuclear site,
\begin{equation}\label{Eq_Phi_gamma}
\Phi^{\text{di},\gamma}_{\kappa\alpha,\kappa'\beta}=
\frac{\partial}{\partial\tau_{\kappa\alpha}}
\mathcal{M}_{\gamma,\kappa'\beta}-
\frac{\partial}{\partial\tau_{\kappa'\beta}}
\mathcal{M}_{\gamma,\kappa\alpha}.
\end{equation}
This additional term was neglected in earlier studies; its explicit calculation
constitutes one of the main technical advances of this work. 
Finally, we have a third contribution in the form
\begin{equation}
\begin{split}
\Phi^{\rm ea,\gamma}_{\kappa\alpha,\kappa'\beta}=&
\frac{1}{2}\sum_{j,l}\epsilon^{\gamma j l}\Big(
\tau_{\kappa' j}\frac{\partial J^{(0)}_{l,\kappa'\beta}}
{\partial\tau_{\kappa\alpha}}-
\tau_{\kappa j}\frac{\partial J^{(0)}_{l,\kappa\alpha}}
{\partial\tau_{\kappa'\beta}}
\Big),
\end{split}
\end{equation}
which is different from zero only when $\kappa\neq\kappa'$,
and corresponds to the \textit{electrical anharmonicity} (ea) tensor discussed by Roman \textit{et al.} 
\cite{PhysRevB.74.245204}. 
This term is present only if the
site symmetries of the occupied Wyckoff position lack the space inversion operation;  
if, on the other hand, every atom in the crystal sits at an inversion center 
(e.g., cubic perovskites like SrTiO$_3$), $\Phi^{\rm ea,\gamma}_{\kappa\alpha,\kappa'\beta}$
vanishes identically.

One can verify that all three contributions are antisymmetric under
$\kappa\alpha\leftrightarrow\kappa'\beta$, consistent with the definition
of Eq.~(\ref{Eq_Phi_gamma_1}) and also that they are independent of the choice of
the coordinate origin.
As a final comment, we expect all these three terms to vanish for large interatomic
distances, although there may be long-range contributions mediated by
electrostatic forces; their detailed analysis, while interesting, goes
beyond the scope of our work, as we will only focus on zone-center phonons.
\subsection{Phonon $g$ factors and frequency splitting}
We now demonstrate how the formalism of  Sec.~\ref{Sec_circ_phonon} can be used to calculate the 
$g$ factor for the phonon modes of the system \cite{PhysRevMaterials.1.014401,PhysRevMaterials.3.064405,ceresoli2002berry}.
Recalling the equations of motion of the ions given by 
Eq.~(\ref{Eq_motion_Phi}), as usual, we seek a solution of the type
\begin{equation}
\label{Eq_phonon}
u^l_{\kappa\beta}(t)=U_{\kappa\beta}^{\mathbf{q}}e^{i(\textbf{q}\cdot\textbf{R}^l_{\kappa}-\omega t)},
\end{equation}
where $\omega$ is the frequency.
We shall specialize to the $\mathbf{q=0}$ case henceforth, and thus remove the $\mathbf{q}$ subscript. 
We obtain,
\begin{equation}\label{Eq_motion}
\begin{split}
\omega^2 \tilde{U}_{\kappa\alpha}=\sum_{\kappa'\beta} \Big(D^{(0)}_{\kappa\alpha,\kappa'\beta}+i\omega B_\gamma D^\gamma_{\kappa\alpha,\kappa'\beta}\Big) \tilde{U}_{\kappa'\beta},
\end{split}
\end{equation}
with $U_{\kappa\alpha}=\tilde{U}_{\kappa\alpha}/\sqrt{M_\kappa}$ and
\begin{equation}\label{Eq_Phi_D_mass}
\begin{split}
D^{(0)}_{\kappa\alpha,\kappa'\beta}&=\frac{1}{\sqrt{M_\kappa M_{\kappa'}}}\sum_l \Phi_{\kappa\alpha,\kappa'\beta}(0,l),\\
D^\gamma_{\kappa\alpha,\kappa'\beta}&=\frac{1}{\sqrt{M_\kappa M_{\kappa'}}}\sum_l \Phi^\gamma_{\kappa\alpha,\kappa\beta}(0,l).
\end{split}
\end{equation}
We shall treat the frequency- and ${\bf B}$-dependent contribution of $D^\gamma_{\kappa\alpha,\kappa'\beta}$
to Eq.~(\ref{Eq_motion}) as a small perturbation of the zero-${\bf B}$ phonon dynamics in the following.

Consider a cubic crystal with a two-fold degenerate transverse optical mode at the $\Gamma$ point 
(e.g., the ``soft'' \cite{PhysRev.134.A981} 
polar mode in cubic SrTiO$_3$).
The unperturbed (zero-${\bf B}$) frequency $\omega^{(0)}$ can be determined by solving the 
following eigenvalue problem
\begin{equation}\label{Eq_unperturbed_eigen}
[\omega_i^{(0)}]^2V^{(i)}_{\kappa\alpha}=\sum_{\kappa'\beta}D^{(0)}_{\kappa\alpha,\kappa'\beta}
V^{(i)}_{\kappa'\beta},
\end{equation}
where $i$ runs over the degenerate modes and $V^{(i)}_{\kappa'\beta}$ are the eigenvector components, where $\kappa'$ runs from 1 to $N$ (number of ions in the cell) and $\beta$ runs over the Cartesian directions.
We choose $i=1,2$ to
span the plane orthogonal to $\mathbf{B}$ in such a way that they form a right handed coordinate system.
We shall now apply degenerate perturbation theory 
to Eq.~(\ref{Eq_unperturbed_eigen}) by choosing the unperturbed 
eigenvectors as 
\begin{equation}\label{Eq_+-}
\begin{split}
\ket*{+}&=\frac{1}{\sqrt{2}}\Big( \ket*{V^{(1)}}+i\ket*{V^{(2)}}  \Big),\\
\ket*{-}&=\frac{1}{\sqrt{2}}\Big( \ket*{V^{(1)}}-i\ket*{V^{(2)}}  \Big),
\end{split}
\end{equation}
where $\bra*{\kappa'\beta}\ket*{V^{(i)}}=V^{(i)}_{\kappa'\beta}$. Here $\ket*{\kappa'\beta}$
stands for a unit displacement of ion $\kappa'$ along the Cartesian direction $\beta$ while the rest of ions remain still; $\ket*{V^{(i)}}$ is therefore a $3\times N$ dimensional vector. 
$\ket*{+}$ and $\ket*{-}$ are circularly polarized phonon modes expressed as a superposition of 
linearly polarized modes.  
In order to account for the frequency splitting and to verify that the eigenvectors given by Eq.~(\ref{Eq_+-}) diagonalize the perturbation, we build the perturbation matrix $g_{ij}$,
\begin{equation}
g_{ij}=i\begin{pmatrix}
\bra*{+}D^\gamma\ket*{+} & \bra*{+}D^\gamma\ket*{-} \\
\bra*{-}D^\gamma\ket*{+} & \bra*{-}D^\gamma\ket*{-} 
\end{pmatrix},
\end{equation} 
which we identify with the gyromagnetic $g_{ij}$ tensor of the phonon modes \cite{PhysRevLett.89.116402,ceresoli2002berry,PhysRevMaterials.1.014401,PhysRevMaterials.3.064405}.
Assuming cubic symmetry, this reduces to 
\begin{equation}
g_{ij}=\begin{pmatrix}
g & 0\\
0 & -g
\end{pmatrix},
\end{equation}
where 
\begin{equation}\label{Eq_g}
\begin{split}
g&=i\bra*{+}D^\gamma\ket*{+}\\
&=i\bra{+}(D^{\text{pc},\gamma}+D^{\text{di},\gamma} )\ket*{+}\\
&=g^\text{pc}+g^\text{di}
\end{split}
\end{equation}
is the $g$ factor of the phonon modes.
We have explicitly indicated the two contributions on $D^{\gamma}$ coming from 
Eq. (\ref{pc}) and Eq. (\ref{Eq_Phi_gamma}); there is only a difference of a mass factor between $\Phi^\gamma$ and $D^\gamma$, which is given in Eq. (\ref{Eq_Phi_D_mass}).
Once the $g$-factor is computed it is easy to give an expression for the frequency splitting of the modes,
\begin{equation}
\omega^{(\pm)}\simeq \omega^{(0)}\pm \frac{1}{2}gB_\gamma.
\end{equation}

Before closing this Section, we briefly comment on the relationship between our methodology to calculate the phonon $g$ factors and previous first-principles approaches. Spaldin and coworkers \cite{PhysRevMaterials.1.014401,PhysRevMaterials.3.064405} calculated the ``pc'' contribution, while the ``di'' term was systematically neglected, resulting in a point-charge approximation to the full $g$ factor; we will show below that for the soft polar mode in SrTiO$_3$, both terms are the same order of magnitude. In Ref. ~\onlinecite{ceresoli2002berry}, Ceresoli presents a point charge model, in addition to a similar perturbative treatment to our Eq. (\ref{Eq_motion}). In the latter, it was assumed that the Born effective charge tensor was isotropic for each sublattice $\kappa$, which is not the case for cubic perovskites like SrTiO$_3$. Also, Ceresoli's version of our dispersion contribution $D^{\text{di},\gamma}$ was in the form of a Berry curvature. While formally equivalent to our expression [which can be seen by writing the Lagrangian in terms of the effective vector potential given by Eq.~(\ref{Eff_vec_pot})], it is more computationally demanding compared to the DFPT implementation given here.
\section{Implementation}\label{Sec_implementation}

We now discuss the practical calculation of the dynamical magnetic moments, $\bm{\mathcal{M}}$,
in the framework of density-functional perturbation theory. (The other materials-dependent 
quantity entering the $g$ factors, i.e., the Born effective charge tensor ${\bf Z}$, is 
straightforward to calculate within standard implementations of DFPT  \cite{PhysRevB.55.10355,PhysRevB.43.7231,PhysRevB.58.6224}.)

\subsection{Polarization response to a long-wavelength phonon}\label{Sec_implementation_A}
As a first step, we 
express the real-space moments of Eq.~(\ref{Eq_P}) in a form that 
is more practical from the computational perspective. To that end, we consider
the macroscopic (cell-averaged) adiabatic current that is associated with the distortion pattern of Eq.~(\ref{Eq_phonon}),
\begin{equation}
\frac{\partial {\bf J}^{\rm mac}({\bf r})}{\partial U_{\kappa\beta}^{\mathbf{q}}} = -i \omega  {\bf P}^{\bf q}_{\kappa\beta} 
 e^{i(\textbf{q} \cdot \textbf{r}-\omega t)}.
\end{equation}
The quantities defined in Eq. (\ref{Eq_P_series}) can then be recast as a long-wave expansion \cite{Stengel2013a},
\begin{equation}
{\bf J}^{\bf q}_{\kappa\beta} = {\bf J}^{(0)}_{\kappa\beta}  -iq_\gamma {\bf J}^{(1,\gamma)}_{\kappa\beta} + \cdots,
\end{equation}
where ${\bf J}^{\bf q}_{\kappa\beta} = \Omega {\bf P}^{\bf q}_{\kappa\beta}$, $\Omega$ being the cell volume.
The advantage of this reciprocal-space formulation is that the macroscopic polarization
response at any ${\bf q}$ can be defined and calculated using a primitive unit cell as 
\begin{equation}
\label{pq}
J^{\bf q}_{\alpha,\kappa\beta} = 
 -2 
  \, {\rm Im}  \int [d^3 k] \sum_n \langle u^{A_\alpha}_{n\bf k,q} | u^{\tau_{\kappa \beta}}_{n\bf k,q}  \rangle,
\end{equation}  
where $\int [d^3k]\equiv \frac{\Omega}{(2\pi)^3}\int_\text{BZ}d^3k$.
Here $| u^{\lambda}_{n\bf k,q} \rangle$ indicates the adiabatic first-order response of the
electronic band $n{\bf k}$ to the perturbation $\lambda$, where $\lambda=A_\alpha$
stands for the electromagnetic vector potential and $\lambda=\tau_{\kappa \beta}$ 
refers to the phonon perturbation Eq.~(\ref{Eq_phonon}) taken at $\omega=0$.
The implementation described in Ref. ~\onlinecite{Dreyer2018} allows one to calculate ${\bf J}^{\bf q}_{\kappa\beta}$
directly via Eq.~(\ref{pq});
$J^{(1,\gamma)}_{\alpha,\kappa\beta}$ can be then obtained by taking numerical derivatives 
around $\textbf{q}=0$.
The same finite-${\bf q}$ implementation~\cite{Dreyer2018} allows one to compute the 
magnetic susceptibility of the system, which we shall use in our numerical tests of 
Eq.~(\ref{Eq_mz2}).

\subsection{Analytical long-wave expansion}\label{Sec_implementation_B}

An alternative approach, which we shall prefer in the context of this work, 
consists in taking the long-wave expansion of Eq.~(\ref{pq}) analytically by
using the formalism described in Ref. ~\onlinecite{Royo_Stengel_PRX}.
A straightforward differentiation of Eq.~(\ref{pq}) leads to
\begin{equation}
\begin{split}
\label{p1}
J^{(1,\gamma)}_{\alpha,\kappa\beta} &= 
 -2\, {\rm Im}  \int [d^3 k] \sum_n  \Big( \langle u^{A_\alpha}_{n\bf k,\gamma} | u^{\tau_{\kappa \beta}}_{n\bf k}  \rangle 
  + \langle u^{A_\alpha}_{n\bf k} | u^{\tau_{\kappa \beta}}_{n\bf k,\gamma}  \rangle \Big),
\end{split}
\end{equation}  
where we have defined the wavefunction response to the spatial gradient of the
perturbation $\lambda$ as
\begin{equation}
| u^{\lambda}_{n\bf k,\gamma}  \rangle = \frac{\partial  | u^{\lambda}_{n\bf k,q} \rangle }{\partial q_\gamma}\Bigg|_{\textbf{q}=0}.
  \end{equation}
Explicit computation of $| u^{\tau_{\kappa \beta}}_{n\bf k,\gamma}  \rangle$ would 
imply a major computational effort; this can be, however, circumvented via a
careful use of the ``$2n+1$'' theorem, \cite{Royo_Stengel_PRX} which yields 
the
second term in the round brackets of Eq.~(\ref{p1}).
The calculation of $\langle u^{A_\alpha}_{n\bf k,\gamma} |$ (first term in
Eq.~(\ref{p1}), indicated as ``T5'' in Ref. ~\onlinecite{Royo_Stengel_PRX})
is comparatively uncomplicated, and can be, in principle, carried out by 
following the guidelines of Ref. ~\onlinecite{Royo_Stengel_PRX}. 
The existing implementation \cite{Royo_Stengel_PRX}, however, focuses on the dynamical quadrupoles 
$Q^{(2,\alpha\gamma)}_{\kappa\beta}$ [see Eq.~(\ref{Eq_Quadrupole})], which are symmetric under exchange of 
Cartesian indices $\alpha\leftrightarrow\gamma$. 
Thus only the \emph{symmetric} components of $\langle u^{A_\alpha}_{n\bf k,\gamma} |$ 
are currently available. To access the \emph{antisymmetric} components, as required by Eq. (\ref{p1}),
the calculation of $\langle u^{A_\alpha}_{n\bf k,\gamma} |$ needs to be generalized as
we describe in the following.
\subsection{Response to a long-wavelength vector potential field}
This section is devoted to give explicit expressions for the response to a vector potential $\mathbf{A}$. A detailed derivation
of the perturbing operators for the response of a vector potential $\mathbf{A}$ is given in the Appendix of Ref. ~\onlinecite{Royo_Stengel_PRX}; here we summarize the main results.
(In general, the response functions have both valence and conduction band components. However, in the present case the valence band part turns out to be irrelevant since it is multiplied by a conduction band state; we focus on the conduction band part in the following.) The wave-function response (in the long-wave limit) to a vector potential can be written in terms of the following Sternheimer equation \cite{Royo_Stengel_PRX}
\begin{equation}\label{Eq_Stern_O}
(\hat{H}^{(0)}_\mathbf{k}+a\hat{P}_\mathbf{k}-\epsilon^{(0)}_{m\mathbf{k}})
\ket*{u^{A_\alpha}_{m\mathbf{k},\gamma}}=-\hat{Q}_\mathbf{k}
\hat{O}_\mathbf{k}\ket*{u^{(0)}_{m\mathbf{k}}},
\end{equation}
where $H^{(0)}_\mathbf{k}$ is the ground state Hamiltonian of the system, $\epsilon_{m\mathbf{k}}^{(0)}$ is its energy eigenvalue, $a$ is a parameter that ensures stability \cite{RevModPhys.73.515} and $\hat{Q}_\mathbf{k}=1-\hat{P}_\mathbf{k}$ is the conduction band projector with $\hat{P}_\mathbf{k}=\sum_m \ket*{u^{(0)}_{m\mathbf{k}}}\bra*{u^{(0)}_{m\mathbf{k}}}$. The perturbing operator $\hat{O}_\mathbf{k}$ in Eq. (\ref{Eq_Stern_O}) is given by
\begin{equation}
\hat{O}_\mathbf{k}=
2\partial_\gamma\hat{H}^{(0)}_\mathbf{k}\partial_\alpha\hat{P}_\mathbf{k}
-2\partial_\gamma\hat{P}_\mathbf{k}\partial_\alpha\hat{H}^{(0)}_\mathbf{k}
+\partial^2_{\gamma\alpha}\hat{H}^{(0)}_\mathbf{k},
\end{equation}
where $\partial_\alpha\equiv \partial/\partial k_\alpha$ is a gradient in $\mathbf{k}$ space. Interestingly, the \textit{symmetric} (S) part (under the exchange $\alpha\leftrightarrow\gamma$) of the perturbing operator $\hat{O}_\mathbf{k}$ corresponds to the $d^2/d\mathbf{k}d\mathbf{k}$ perturbation,
\begin{equation}
\begin{split}
\hat{O}_\mathbf{k}^S=&
+\partial_\gamma\hat{H}^{(0)}_\mathbf{k}\partial_\alpha\hat{P}_\mathbf{k}
+\partial_\alpha\hat{H}^{(0)}_\mathbf{k}\partial_\gamma\hat{P}_\mathbf{k}\\
&-\partial_\gamma\hat{P}_\mathbf{k}\partial_\alpha\hat{H}^{(0)}_\mathbf{k}
-\partial_\alpha\hat{P}_\mathbf{k}\partial_\gamma\hat{H}^{(0)}_\mathbf{k}
+\partial^2_{\gamma\alpha}\hat{H}^{(0)}_\mathbf{k},
\end{split}
\end{equation}
which is already implemented in the publicly available {\sc abinit} code \cite{gonze2009abinit,gonze2020abinit}; while its \textit{antisymmetric} (A)
contributions gives rise to the response of a uniform $\mathbf{B}$ field, as defined in 
Ref. ~\onlinecite{PhysRevB.81.205104},
\begin{equation}
\begin{split}
\hat{O}_\mathbf{k}^A=&
+\partial_\gamma\hat{H}^{(0)}_\mathbf{k}\partial_\alpha\hat{P}_\mathbf{k}
-\partial_\alpha\hat{H}^{(0)}_\mathbf{k}\partial_\gamma\hat{P}_\mathbf{k}\\
&-\partial_\gamma\hat{P}_\mathbf{k}\partial_\alpha\hat{H}^{(0)}_\mathbf{k}
+\partial_\alpha\hat{P}_\mathbf{k}\partial_\gamma\hat{H}^{(0)}_\mathbf{k}.
\end{split}
\end{equation}
We therefore conclude that the computational cost of calculating the response to a vector potential as defined in Eq.~(\ref{Eq_Stern_O}) and the response to a uniform $\mathbf{B}$ field is the same as for the usual $d^2/d\mathbf{k}d\mathbf{k}$ perturbation. Furthermore, given the similarities of the perturbing operators (their differ only by a couple of signs) their implementation turns out to be straightforward. 

\subsection{Computational parameters}
The formalism described in Secs.~\ref{Sec_implementation_A} and \ref{Sec_implementation_B} has been implemented in the {\sc abinit} code \cite{gonze2009abinit,gonze2020abinit,PhysRevB.55.10355,PhysRevB.55.10337}.  
We use the Perdew-Wang \cite{PhysRevB.45.13244} parametrization of the local density approximation (LDA) and
Optimized Norm-Conserving Vanderbilt Pseudopotentials (ONCVPSP) \cite{PhysRevB.88.085117} in all the DFT and DFPT calculations.

Our numerical results on rotational $g$ factors in molecules are obtained
employing 
a large cell of $20\cross20\cross20$ 
bohr$^3$ to avoid interactions between neighboring images. A maximum plane-wave cutoff of $100$ Ha (60 Ha for CH$_4$, C$_5$H$_5$N and C$_6$H$_5$F) is used 
and the Brillouin zone is sampled with a single $\mathbf{k}$ point at $\Gamma$. 
The structural optimization for the geometry of the molecules is performed 
to a tolerance of $5\cdot10^{-7}$ Ha/bohr on the residual forces.
 
For our calculations on SrTiO$_3$,
we use the five-atom primitive cubic cell, with a plane-wave cutoff of 80 Ha and an $8\cross8\cross8$ mesh of 
$\mathbf{k}$ points to sample the Brillouin zone;
with this setup we obtain an optimized cell parameter of
$a_0=7.288$ bohr. 
For the derivative with respect to the displacement of atoms appearing in Eq. (\ref{Eq_Phi_gamma}), $\partial/\partial\tau_{\kappa\alpha}$,
we apply a displacement of 0.01 bohr to atom $\kappa$ along the Cartesian direction $\alpha$ and compute the derivative via finite differences; this means that $3N$ (where $N$ is the number of atoms in the cell) of such calculations are needed to compute the full $D^{\gamma}$ matrix.
This number could be reduced significantly via use of symmetries; however, in our calculations we opt
for a straightforward calculation of all components, and check that the resulting generalized Lorentz force
tensor enjoys the expected symmetries as part of the validation procedure.

\section{Results}\label{Sec_Results}
\subsection{Rotational $g$ factor of molecules} \label{Sec_Res_g}
To begin with, we present a detailed study of the H$_2$, N$_2$, and F$_2$ molecules, 
since they constitute the simplest nontrivial test of our methodology.
In the case of elemental diatomic molecules, the gyromagnetic $g$ factor is only defined for 
rotations about an axis that is perpendicular to the bond.
Assuming that the bond is aligned with the $x$ Cartesian direction,
and that the rotation axis passes through the center of mass,
the $g$ factor reduces to 
\begin{equation}
\label{eq_dia}
g=\frac{J^{(1,x)}_{y,1y}-J^{(1,y)}_{x,1y}}{I},
\end{equation} 
where $I=Md^2/2$ is the moment of inertia. ($d$ stands for
the interatomic distance, and $M$ is the atomic mass in units of the proton mass.)

\begin{figure}
	\includegraphics[width=1.0\linewidth]{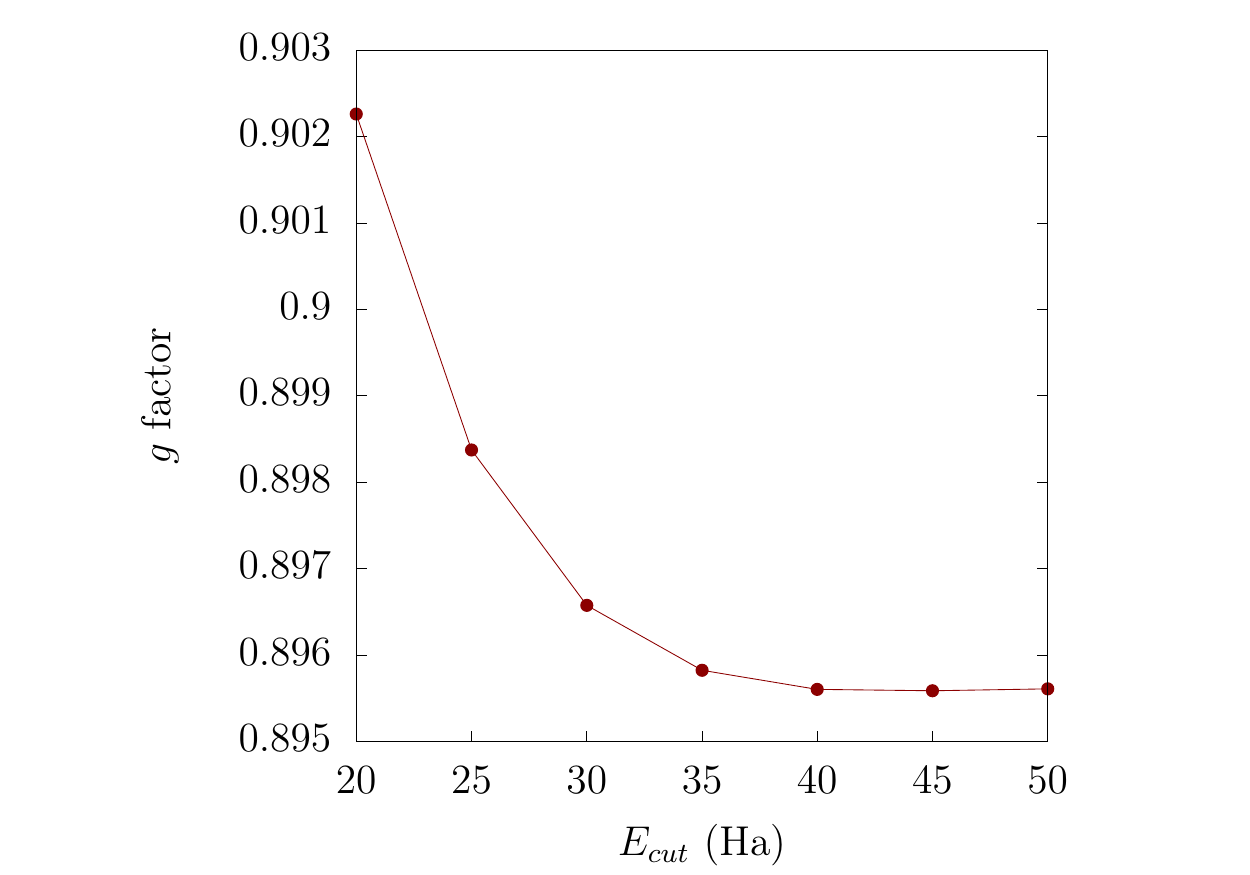}
	\caption{Convergence of the $g$ factor of H$_2$ (with $d_\text{exp}=1.4$ bohr) with respect to the plane-wave cutoff. Calculations are performed using a single $\mathbf{k}$ point ($\Gamma$) with a box size of $20^3$ bohr$^3$.}
	\label{Fig_H2_ecut}
\end{figure}

Figure \ref{Fig_H2_ecut} shows the 
convergence with the plane-wave cutoff of the $g$ factor of
H$_2$ using the experimental geometry ($d_\text{exp}$=1.4 bohr), calculated using the analytical 
long-wave approach described in Sec.~\ref{Sec_implementation_B}.
We see that the result is well-converged at a relatively modest (for a molecule in a box) cutoff of 50 Ha. 
We can compare the converged value of 0.8956 to the finite-$\textbf{q}$ calculations described in 
Sec.~\ref{Sec_implementation_A}, which gives precisely 0.8956.
For $N_2$ ($d_\text{exp}$=2.074 bohr) and $F_2$ ($d_\text{exp}$=2.668 bohr), the analytical long-wave approach gives  
$-0.2704$ and $-0.1043$, also in excellent agreement with the finite-difference method, which yields 
$-0.2708$ and $-0.1045$, respectively. 
The excellent agreement confirms the accuracy of 
our implementation described in Sec.~\ref{Sec_implementation_B}.

\begin{figure}
	\includegraphics[width=1.0\linewidth]{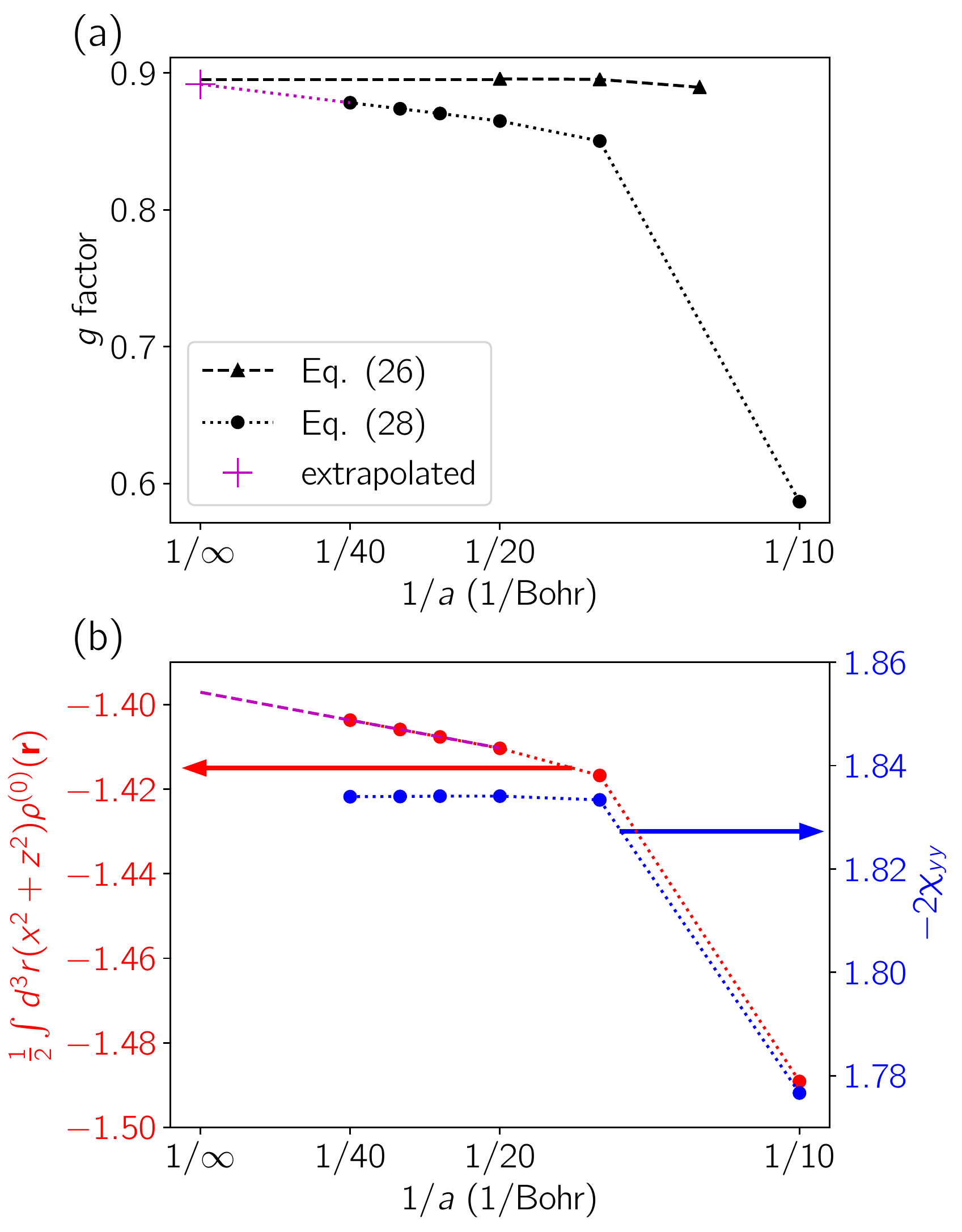}
	\caption{(a) Calculation of rotational $g$ factor of H$_2$ (with $d_\text{exp}$=1.4 bohr) using the expression for $\partial m_a/\partial\omega_b$ from Eq.~(\ref{Eq_mz2}) (dots) and from Eq. (\ref{Eq_mz}) (triangles) versus inverse of the simulation cell size side length. (b) Convergence of terms in Eq.~(\ref{Eq_mz2}) versus inverse of the simulation cell size. Purple dashed line in (b) is the extrapolated value for the quadrupole term; the purple cross in (a) is the $g$ factor calculated with the extrapolated quadrupole term.}
	\label{Fig_alt_H2}
\end{figure}

Since the H atom is well described by a local pseudopotential, we can use the H$_2$ molecule to benchmark the performance of the 
two alternative formulations of $\partial m_a/\partial\omega_b$, i.e. Eq.~(\ref{Eq_mz}) [which reduces to Eq.~(\ref{eq_dia})
in this case] and Eq.~(\ref{Eq_mz2}). 
In Fig.~\ref{Fig_alt_H2}(a) we the plot the calculated $g$ factor for H$_2$ versus inverse cell size by using either method.
As we anticipated in Sec. \ref{Sec_mag_rotation}, we find that Eq.~(\ref{Eq_mz2}) is quite challenging to converge, while
the corresponding results of Eq.~(\ref{eq_dia}) display an optimally fast convergence.
To understand the origin of such a behavior, we show in Fig.~\ref{Fig_alt_H2}(b) a decomposition of Eq.~(\ref{Eq_mz2})
 into the two contributions on the rhs.
 This analysis clarifies that the convergence of  is limited by the quadrupole term 
 [i.e., the second term in Eq.~(\ref{Eq_mz2})], while the magnetic susceptibility of the molecule 
 is already converged at a relatively small box size. 
 If we extrapolate this term to the limit of an infinitely large cell parameter ($1/a\rightarrow 0$, purple dashed curve), then we see that our $g$ factor indeed converges to the value we obtain using the methodology of Sec.~\ref{Sec_implementation_B} [purple cross on  Fig.~\ref{Fig_alt_H2}(a)]. The agreement for large cell sizes provides an independent confirmation of the accuracy of our approach, though the methodology of  Sec.~\ref{Sec_implementation_B} is clearly superior from a computational perspective.
 
As we anticipated, a further issue with Eq.~(\ref{Eq_mz2}) consists in the fact that it may yield qualitatively 
 incorrect results when nonlocal pseudopotentials are used, i.e., in the vast majority of first-principles simulations
 that are being performed nowadays.
 An obvious example is that of a neutral (and isolated) closed-shell atom, where the rotationally induced magnetization must vanish 
 exactly. 
 This requirement is trivially fulfilled by our Eq.~(\ref{Eq_mz}): both dynamical charges and dynamical magnetic 
 moments identically vanish in this system due to charge neutrality and inversion symmetry.
 In the context of Eq.~(\ref{Eq_mz2}) one would expect a vanishing result, too: Langevin's theory of diamagnetism
 expresses the susceptibility as the quadrupolar moment of the spherical atomic charge, which should cancel out exactly 
 with the second term on the rhs.
 In presence of nonlocal pseudopotentials, however, Langevin's result no longer holds, and Eq.~(\ref{Eq_mz2})
 yields a nonzero value for all noble gas atoms except He. (The latter, just like H, is well 
 described by a local pseudopotential.) We regard this as a serious concern in this context, and we therefore caution against 
 a straightforward application of Eq.~(\ref{Eq_mz2}) to the calculation of rotational $g$ factors.

In addition to the aforementioned elemental diatomic molecules, we consider 
several other examples:
HF, HNC, and FCCH (still linear, but with a finite dipole moment),  
nonlinear molecules such as NH$_3$, H$_2$O, and CH$_4$, and the aromatic compounds C$_5$H$_5$F and C$_6$H$_5$F. 
At difference with H$_2$ and related structures, in all these cases Eq.~(\ref{Eq_gfac}) contains a nonzero 
contribution from the Born effective charges; therefore, these additional examples 
provide us with the opportunity to test the full formula, Eq.~(\ref{Eq_mz}) 
[in combination with Eq. (\ref{Eq_gfac})], rather than its simplified version, Eq.~(\ref{eq_dia}).
The molecular geometries and rotational axes used in this work 
are discussed in Appendix \ref{structure}.

\begin{table}
  	\caption{Calculated rotational $g$ factors for selected simple molecules compared with the relevant literature data. 
  	``HF/DFT'' and ``MP/CCD'' stand for computational results at various level of theory (Hartree-Fock / Density Functional Theory, 
  	and M\o{}ller-Plesset perturbation theory / Coupled-Cluster with Double excitations); ``Exp.'' refers to experimental measurements.}
	\begin{ruledtabular}
          \label{tab:molecule_g}
          \begin{tabular}{c|rrrr}
            &\multicolumn{4}{c}{Rotational $g$ factor} \\
			&This work&  HF/DFT &MP/CDD& Exp.  \\ 
			\hline
			\multirow{2}*{H$_2$}& \multirow{2}*{0.8901}  & $0.9103$\footnotemark[1] &\multirow{2}*{0.8899\footnotemark[1]}&\multirow{2}*{$0.8829$\footnotemark[3]} \\
			&&0.8755\footnotemark[2]&&\\\hline
			N$_2$& $-0.2699$ &$-0.2872$\footnotemark[1] &$-0.2653$\footnotemark[1]&$-0.2593$\footnotemark[3]  \\\hline
			F$_2$& $-0.1003$  & $-0.0900$\footnotemark[1] &$-0.1136$\footnotemark[1]&$-0.1208$\footnotemark[3]\\\hline
			HF & 0.7603 &0.7624\footnotemark[1] &0.7488\footnotemark[1]&0.7392\footnotemark[3]\\\hline
		    HNC & $-0.1004$  &$-0.0996$\footnotemark[1] &$-0.0968$\footnotemark[1]&\\\hline
		    FCCH&$-0.0065$ &&&$-0.0077$\footnotemark[4]\\\hline
		    H$_2$O& 0.6699 &0.6640\footnotemark[1]&0.6507\footnotemark[1]&0.6450\footnotemark[3]\\\hline
		    NH$_3$& 0.5289  &0.5061\footnotemark[1]&0.5044\footnotemark[1]&\\\hline
		    \multirow{2}*{CH$_4$} & \multirow{2}*{0.3629}   &0.3019\footnotemark[1] &\multirow{2}*{0.3190\footnotemark[1]}&\multirow{2}*{$0.3133$\footnotemark[3]}\\
			&&0.2985\footnotemark[5]&\\\hline
		    C$_5$H$_5$N& 0.0411 &&&0.0428\footnotemark[4]\\\hline
		    C$_6$H$_5$F& 0.0276&&&0.0266\footnotemark[4]\\\hline
		\end{tabular}
	\end{ruledtabular}
              \footnotetext[1]{Ref. ~\onlinecite{cybulski1997calculation}}
              \footnotetext[2]{Ref. ~\onlinecite{PhysRevLett.89.116402}}
              \footnotetext[3]{Ref. ~\onlinecite{doi:10.1080/00268977100100221}}
              \footnotetext[4]{Ref. ~\onlinecite{flygare1974magnetic}}
              \footnotetext[5]{Ref. ~\onlinecite{C_T_private}}
            \end{table}

In Table \ref{tab:molecule_g} we compare our results for the rotational $g$ factors 
to experimental measurements from Refs. ~\onlinecite{doi:10.1080/00268977100100221} and ~\onlinecite{flygare1974magnetic}.
In addition, we report the results of previous calculations using 
Hartree-Fock (HF) and post Hartree-Fock methods\cite{cybulski1997calculation}, 
as well as DFT calculations using the Berry-phase method \cite{PhysRevLett.89.116402,C_T_private}. 
Since the inclusion of electron-electron 
correlations, either at at the level of M\o{}ller-Plesset (MP) perturbation theory or coupled cluster with 
double excitations (CCD), seems to improve the agreement with experiment in many cases,~\cite{cybulski1997calculation} 
we include those data as well for comparison.  
We see that our DFPT based method compares well even with the best theoretical values obtained via more computationally demanding methods.
Our results in Table~\ref{tab:molecule_g} are also in excellent agreement with experiment, where available. 
CH$_4$ appears to be the only exception, though the reason for the larger  discrepancy is not clear. 
\subsection{Soft-mode frequency splitting of cubic SrTiO$_3$}
\begin{table}
  	\caption{$g$ factor of the soft polar TO mode at the zone center in cubic SrTiO$_3$. Units are in 
	$10^{-4}$  atomic units.
		\label{tab:STO}}
	\begin{ruledtabular}
		\begin{tabular}{cccc}
		     & $g$ & $g^\text{pc}$ & $g^\text{di}$  \\ 
			\hline
		  SrTiO$_3$	& $-1.2083$ & $0.6679$  & $-1.8763$ 
		\end{tabular}
	\end{ruledtabular}
\end{table}

We now turn to the splitting of the soft polar TO mode at the zone center in cubic SrTiO$_3$. 
As we did in the case of the rotational $g$ factors in Sec.~\ref{Sec_Res_g}, we can test the accuracy of our generalized Lorentz forces by comparing the implementation described in Sec.~\ref{Sec_implementation_B} with the alternative approach of Sec.~\ref{Sec_implementation_A}. 
In Table~\ref{tab:STO_comp} of Appendix~\ref{App_comp} we present the components of  
$(D^{\text{di,}\gamma})_{\kappa\alpha,\kappa'\beta}$ elements [see Eq.~(\ref{Eq_Phi_gamma})] for cubic SrTiO$_3$ using both methods; we see quite good agreement, giving us confidence that $g^\text{di}$ is accurately calculated.

The results for the $g$ factors are shown in Table \ref{tab:STO}.
Following Eq.~(\ref{Eq_g}), we separate the two different contributions coming from the 
$\mathbf{J}^{(0)}$ ($g^\text{pc}$) and $\mathbf{J}^{(1)}$ ($g^\text{di}$) terms. 
As mentioned earlier, some works\cite{PhysRevMaterials.1.014401,PhysRevMaterials.3.064405} have only accounted for the terms depending on
the Born effective charges within a point-charge approximation, roughly corresponding to our calculated $g^\text{pc}$. 
It is immediately clear from Table \ref{tab:STO} that such an approximation is inappropriate: the remainder ($g^\text{di}$) has opposite sign
and is almost three times larger (in absolute value) than the contribution coming from $g^\text{pc}$; as a result, the total $g$ factor
disagrees with $g^\text{pc}$ both in magnitude and sign.
This indicates that an accurate computation of the $\mathbf{J}^{(1)}$ tensor is crucial in this particular case 
and that these terms should not be neglected. 

For a more quantitative comparison, note that Ref. ~\onlinecite{ceresoli2002berry} and 
Ref. ~\onlinecite{PhysRevMaterials.1.014401}
computed $g^\text{pc}$ for tetragonal SrTiO$_3$, obtaining values of 
$g^\text{pc}=5.76\cdot 10^{-5}\text{cm}^{-1}/\text{T}$ and 
$g^\text{pc}=4.78\cdot 10^{-5}\text{cm}^{-1}/\text{T}$, respectively. 
In those units, our result for cubic SrTiO$_3$ is $g^\text{pc}=6.23\cdot 10^{-5}\text{cm}^{-1}/\text{T}$.
The agreement is rather good, especially considering that: (i) we are considering the full tensorial form of the Born effective charge 
tensor and (ii) our analysis is carried out in the cubic, and not tetragonal, phase of SrTiO$_3$.
Note that Ref. ~\onlinecite{ceresoli2002berry} also reports a result for the total $g$-factor,
$g=-7.95\cdot 10^{-5}\text{cm}^{-1}/\text{T}$, which again compares well to our calculated value of
$g=-11.28\cdot 10^{-5}\text{cm}^{-1}/\text{T}$.

To gain some insight on the physics, we perform a further decomposition of $g^\text{pc}$ and $g^\text{di}$ into 
the individual contributions of each atomic sublattice.
In the case of $g^\text{pc}$, such a decomposition is straightforward, as this term mediates an on-site 
coupling between the displacement of each atom and its own velocity. [This can be appreciated by observing
that the corresponding contribution to the generalized Lorentz force, Eq.~(\ref{pc}), contains
a $\delta_{\kappa \kappa'}$ prefactor.]
The case of $g^\text{di}$ is less obvious: the nondiagonal (on the atomic index) nature of $\Phi^{\rm di}$
implies that the velocity of a given atom can produce forces not only onto itself, but also on its
neighbors.
Thus, prior to attempting a decomposition of $g^\text{di}$, we first isolate the 
basis-diagonal $\kappa=\kappa'$ terms in $\Phi^{\rm di}$, and use them to define 
an on-site contributions to $g^\text{di}$ (indicated as $g^\text{di}_{\kappa=\kappa'}$
henceforth).
Apart from enabling the aforementioned decomposition, this analysis also gives a flavor 
of the overall importance of the off-site contributions to $g^\text{di}$.
\begin{table}
  	\caption{Contribution of each atom to $g^\text{pc}$ and $g^\text{di}_{\kappa=\kappa'}$, which are defined as the on-site ($\kappa=\kappa'$) contributions
  	to $g^\text{di}$. 
  	 Units are in $10^{-4}$ atomic units.
		\label{tab:individual_atom}}
	\begin{ruledtabular}
		\begin{tabular}{ccccccc}
			& Sr & Ti  & O$_1$ &O$_2$&O$_3$& total\\ 
			\hline
			$g^\text{pc}$ & -0.0197 & -0.1082 & 0.2985&0.2985 &0.1987& 0.6679\\
			$g^\text{di}_{\kappa=\kappa'}$ & -0.0118 &-0.1246  & -0.6429 &-0.6429 &-0.0062 &-1.4284
		\end{tabular}
	\end{ruledtabular}
\end{table} 

The results are summarized in Table \ref{tab:individual_atom}. Regarding $g^\text{pc}$, we find that 
the contribution of the oxygen atoms largely dominates over the rest, consistent with the
conclusions of Ref. ~\onlinecite{ceresoli2002berry}. Due to their smaller mass, oxygens evolve along larger orbits,
which amplifies their contribution to the magnetic moment. 
Regarding $g^\text{di}$, we find that the on-site terms represent more than the $75\%$ of the total $g^\text{di}$ factor,
which indicates that intersite couplings have a relatively minor importance. 
At the level of $g^\text{di}_{\kappa=\kappa'}$, we find that the contribution of the 
equatorial oxygens is by far the largest, and primarily responsible for reversing the sign 
of the overall $g$ factor. 
 
Finally, we use the above results to calculate the frequency splitting of the TO modes.  
Considering a magnetic field of $B=100$ T we obtain $gB\sim 0.01\text{ cm}^{-1}$, of the same order
as predicted in Ref. ~\onlinecite{PhysRevMaterials.3.064405}. 
This is a very small value that appears challenging to resolve 
even for the most powerful experimental techniques available nowadays.
Our hope is that the computational tools developed here allow for
a more efficient screening of candidate materials where this effect 
may be measurable.
\section{Conclusions}\label{Sec_conclusions}

We have developed a complete theoretical approach for calculating orbital magnetization 
from rotations and pseudorotations (circularly polarized optical phonons) within the context 
of first-principles theory. 
The approach is based on density-functional perturbation theory calculations of the polarization 
induced by an atomic displacement (i.e., Born effective charges), and its first real-space moment. 
We have demonstrated an implementation to calculate the latter quantity via generalization of the 
existing long-wave approach to dynamical quadrupoles; thus, we have established a connection between 
spatial dispersion phenomena and orbital magnetism, and demonstrated its accuracy via comparison with 
finite-difference calculations. 
Our methodology allows for efficient and optimally accurate computation, and works equally well 
for molecules and solids. 
We have used this approach to determine rotational $g$ factors of some simple molecules, and 
demonstrated excellent agreement with experimental results where available. 
Finally, we have developed a strategy to calculate the generalized Lorentz force on atoms in 
presence of a magnetic field, and utilized it to study the splitting of the soft optical phonons 
in cubic SrTiO$_3$. In the latter system, we demonstrated that contributions to phonon $g$ factor from 
the first moment of the induced polarization, which had been neglected in some previous approaches, 
dominate the response.

In spite of this correction, the overall $g$ factor remains of the same order of magnitude 
as the values quoted in Refs. ~\onlinecite{PhysRevMaterials.3.064405,ceresoli2002berry}. 
Therefore, our theory as it stands appears unlikely to explain the large phonon Hall~\cite{PhysRevLett.124.105901} effects reported experimentally. 
To move forward in this direction, we suspect that it may be necessary to take into account the quantum
paralectric nature of SrTiO$_3$ at low temperatures, e.g., by going beyond the Ehrenfest Lagrangian
of Eq.~(\ref{ehrenfest}).
We regard this as an exciting avenue for further study.

\begin{acknowledgments}
  CED acknowledges support from the National Science Foundation under Grant No.~DMR-1918455. The Flatiron Institute is a
division of the Simons Foundation.

AZ and MS acknowledge support from Ministerio de Economia,
Industria y Competitividad (MINECO-Spain) through
 Grant No. PID2019-108573GB-C22; from Severo Ochoa FUNFUTURE center of excellence (CEX2019-000917-S);
 from Generalitat de Catalunya (Grant No. 2017 SGR1506); and from
 the European Research Council (ERC) under the European Union's
 Horizon 2020 research and innovation program (Grant
 Agreement No. 724529).
\end{acknowledgments}

\appendix

\section{Translational symmetry of the geometric magnetization}

\label{app:shift}

To see how a change in the assumed center of rotation (and, simultaneously,
in the gauge origin) affects the result, consider 
\begin{equation}
\begin{split}
\frac{\partial m_a({\bf R})}{\partial\omega_b} 
=& \sum_{\kappa,j,\beta}\epsilon^{b j\beta}(\tau_{\kappa j} -R_j) \times \\
& \left[ \mathcal{M}_{a,\kappa\beta} +\frac{1}{2} 
 \sum_{i,\alpha}
\epsilon^{ai\alpha} (\tau_{\kappa i}-R_i) J^{(0)}_{\alpha,\kappa\beta}\right] \\
=& -\sum_{\kappa,j,\beta} \epsilon^{b j\beta} R_j \left[ \mathcal{M}_{a,\kappa\beta} 
  + \frac{1}{2} \sum_{i,\alpha}
\epsilon^{ai\alpha} \tau_{\kappa i} J^{(0)}_{\alpha,\kappa\beta}\right] \\
& - \frac{1}{2} \sum_{\kappa,j,\beta,i,\alpha} \epsilon^{b j\beta} \epsilon^{ai\alpha} 
 \tau_{\kappa j} R_i J^{(0)}_{\alpha,\kappa\beta} + \frac{\partial m_a({\bf 0})}{\partial\omega_b}\\
=& - \frac{1}{2} \sum_{\kappa,j,i,\alpha} \epsilon^{b j\alpha} \epsilon^{ai\alpha} R_j \mathcal{D}_i \\
& - \frac{1}{2} \sum_{\kappa,j,\beta,i,\alpha} \epsilon^{b j\beta} \epsilon^{ai\alpha} 
 \tau_{\kappa j} R_i J^{(0)}_{\alpha,\kappa\beta} + \frac{\partial m_a({\bf 0})}{\partial\omega_b}.
\end{split}
\end{equation}
[In the last step we have used the sum rule Eq.~(\ref{summk}).]
It is useful now to observe the following property of the Levi-Civita symbol,
\begin{equation}
\sum_{i,j,\alpha} \epsilon^{b j\alpha} \epsilon^{ai\alpha} = \delta_{ab} \delta_{ij} - \delta_{bi} \delta_{aj}.
\end{equation}
This leads to 
\begin{equation}
\begin{split}
- \frac{1}{2} \sum_{\kappa,j,i,\alpha}  \epsilon^{b j\alpha} \epsilon^{ai\alpha} R_j \mathcal{D}_i 
=& -\frac{1}{2} (\delta_{ab} {\bf R} \cdot \bm{\mathcal{D}} - R_a \mathcal{D}_b).
\end{split}
\end{equation}
In the second line we can write
\begin{equation}
\label{trick}
\begin{split}
-\sum_\kappa \tau_{\kappa j} J^{(0)}_{\alpha,\kappa\beta} =& \sum_\kappa J^{(1,j)}_{\alpha,\kappa\beta} - \mathcal{D}_j.
\end{split}
\end{equation}
The second term yields the same as above, with the $ab$ indices switched; the
first term can be written in terms of $J^{(1)}$. The final result, after collecting all
the contributions is
\begin{equation}
\begin{split}
\frac{\partial m_a({\bf R})}{\partial\omega_b} &- \frac{\partial m_a({\bf 0})}{\partial\omega_b} =\\
 =& -\frac{1}{2} (\delta_{ab} {\bf R} \cdot \bm{\mathcal{D}} - R_a \mathcal{D}_b - R_b \mathcal{D}_a) \\
 & + \frac{1}{2} \sum_{j,\beta,i,\alpha} \epsilon^{b j\beta} \epsilon^{ai\alpha} 
  R_i \sum_\kappa J^{(1,j)}_{\alpha,\kappa \beta}.
\end{split}
\end{equation}
The second term on the rhs vanishes: the sublattice sum of the $J^{(1,j)}_{\alpha,\kappa \beta}$
tensor coincides with the proper piezoelectric tensor times a trivial 
volume factor, and is therefore \emph{symmetric} with respect to $\beta j$. 
(An antisymmetric contribution would describe a steady macroscopic current 
that is generated by a rotating body in its comoving reference frame, and must vanish 
on general physical grounds, see Sec.~III.D.2 of Ref. ~\onlinecite{PhysRevB.98.125133}.)
The remainder leads to Eq.~(\ref{gauge}).

\section{Dipolar sum rule for bounded systems}

\label{app:dipolar}

{\em Statement of the problem.}
We will prove the following sum rule, valid for an isolated molecule in 
open electrostatic boundary conditions,
\begin{equation}
\label{sumrule}
\sum_\kappa \left( J_{\alpha,\kappa\beta}^{(1,\gamma)} + \tau_{\kappa \gamma} Z^*_{\alpha,\kappa \beta}   \right) = 
  \delta_{\alpha \beta} \mathcal{D}_{\gamma},
\end{equation}  
where $\bm{\mathcal{D}}$ is the static dipole moment of the molecule,
\begin{equation}
\mathcal{D}_{\gamma} = \int d^3r r_\gamma \rho^{(0)}({\bf r}).
\end{equation}
In absence of nonlocal pseudopotentials the proof is straightforward: it
suffices to observe that $Z^*_{\alpha,\kappa \beta} = J^{(0)}_{\alpha,\kappa \beta}$,
and then use the definition of the $J^{(n)}_{\alpha,\kappa \beta}$ moments provided in the
main text together with the following relation (translational invariance) for the microscopic
polarization response,
\begin{equation}
\label{translational}
\sum_{\kappa} \mathcal{P}_{\alpha, \kappa \beta}({\bf r}) =  \delta_{\alpha \beta} \rho^{(0)}({\bf r}).
\end{equation}
If nonlocal pseudopotentials are present, Eq.~(\ref{translational}) breaks
down; however, we will show that Eq.~(\ref{sumrule}) is exact even in that case.

{\em Proof.}
To prove Eq.~(\ref{sumrule}) without passing through Eq.~(\ref{translational}),
we will use another (exact) sum rule, relating the $J_{\alpha,\kappa\beta}^{(1,\gamma)}$
moments to the clamped-ion piezoelectric tensor,
\begin{equation}
\label{piezo}
-\frac{1}{\Omega} \sum_\kappa J_{\alpha,\kappa\beta}^{(1,\gamma)} = \bar{e}_{\alpha,\beta \gamma}.
\end{equation}
To apply this rule, we need first of all to place the isolated molecule in a large box of
volume $\Omega$, and work in periodic boundary conditions. 
Then, Eq.~(\ref{piezo}) describes the \emph{proper} piezoelectric response of the resulting
crystal lattice to an infinitesimal strain.
[To avoid complications due to long-range interactions between repeated images, we will
assume that the Coulomb kernel is cut off at the boundary of the box, and that all objects
entering Eq.~(\ref{sumrule}) are consistently calculated in such conditions.]

Since the images of the molecule are isolated in space, the
macroscopic polarization of the crystal is exactly given by the Clausius-Mossotti 
formula as the static dipole moment divided by the volume,
\begin{equation}
{\bf P} = \frac{\bm{\mathcal{D}}}{\Omega}.
\end{equation}
$\bar{e}_{\alpha,\beta \gamma}$, however, is not defined as a straightforward strain
derivative of ${\bf P}$ (that would be the so-called improper piezoelectric tensor).
To arrive at $\bar{e}_{\alpha,\beta \gamma}$ we first need to introduce the
direct lattice vectors ${\bf a}_i$ and their duals ${\bf b}_i$ in such a way that
${\bf a}_i \cdot {\bf b}_j = \delta_{ij}$.
Then, the \emph{reduced} polarization is defined in units of charge as the
flux of ${\bf P}$ through a facet of the crystal cell,
\begin{equation}
p_i = \Omega  {\bf b}_i \cdot {\bf P} = {\bf b}_i \cdot \bm{\mathcal{D}}.
\end{equation}
Finally, the proper piezoelectric tensor is defined as
\begin{equation}
\bar{e}_{\alpha,\beta \gamma} = \frac{1}{\Omega} \sum_i a_{\alpha i} \frac{\partial p_i}{\partial \eta_{\beta \gamma}},
\end{equation}
where $\boldsymbol{\eta}$ is the Cauchy infinitesimal strain tensor.
This leads to the following formula, without factors of volume,
\begin{equation}
\sum_\kappa J_{\alpha,\kappa\beta}^{(1,\gamma)}  =  
-\sum_i ({\bf a_i})_\alpha \frac{\partial ( {\bf b}_i \cdot \bm{\mathcal{D}})}{\partial \eta_{\beta \gamma}}.
\end{equation}

In order to calculate the derivative of the scalar product, note that an infinitesimal strain
corresponds to the following linear transformation of the atomic coordinates and direct lattice vectors,
\begin{subequations}
\begin{align}
\bm{\tau}'_\kappa =& ({\bf I} + \bm{\eta}) \bm{\tau}_\kappa, \\
      {\bf a}'_i  =& ({\bf I} + \bm{\eta}) {\bf a}_i.
\end{align}
\end{subequations}
The first relation yields 
\begin{equation}
\frac{\partial \tau_{\kappa \sigma}}{\partial \eta_{\beta \gamma}} = \delta_{\beta \sigma} \tau_{\kappa \gamma},
\end{equation}
and immediately (by using the definition of the Born charge tensor),
\begin{equation}
\frac{\partial \mathcal{D}_\alpha}{\partial \eta_{\beta \gamma}} = 
  \sum_{\kappa \sigma} \frac{\partial \mathcal{D}_\alpha}{\partial \tau_{\kappa \sigma}} 
  \frac{\partial \tau_{\kappa \sigma}}{\partial \eta_{\beta \gamma}} =
 \sum_\kappa Z^*_{\alpha,\kappa \beta} \tau_{\kappa \gamma}.
\end{equation}  
The second relation is used to determine the transformation law for the duals.
The reciprocal-space vectors need to preserve the orthonormality condition to 
linear order in the strain, which leads to the following result,
\begin{equation}
{\bf b}'_i  \simeq ({\bf I} - \bm{\eta}^{T}) {\bf b}_i.
\end{equation}
From this, we deduce 
\begin{equation}
\frac{\partial ({\bf b}_i)_\xi}{\partial \eta_{\beta \gamma}} = -\delta_{\gamma \xi} ({\bf b}_i)_\beta.
\end{equation}
By using the orthonormality rule $\sum_i ({\bf a_i})_\alpha ({\bf b}_i)_\beta = \delta_{\alpha \beta}$,
we eventually arrive at
\begin{equation}
\sum_\kappa J_{\alpha,\kappa\beta}^{(1,\gamma)}  =  
\mathcal{D}_\gamma - \sum_\kappa Z^*_{\alpha,\kappa \beta} \tau_{\kappa \gamma},
\end{equation}
thereby concluding our proof.

{\em General consequences.}
The above results allow us to further specify the validity of Eq.~(\ref{translational})
in the case of an isolated molecule. While the microscopic formula breaks down in
presence of nonlocal pseudopotentials, one can expand both sides into Cartesian multipoles 
and ask at what order the equality no longer holds.
At order zero the equality clearly holds,
\begin{equation}
\label{tra0}
\sum_{\kappa} \int d^3 r \mathcal{P}_{\alpha, \kappa \beta}({\bf r}) =  \delta_{\alpha \beta} \int d^3r \rho^{(0)}({\bf r}),
\end{equation}
since macroscopic currents are well described; in the case of a neutral 
molecule Eq.~(\ref{tra0}) reduces to the acoustic sum rule on the Born charge tensor components.
In this Appendix, we have provided a formal proof that Eq.~(\ref{translational}) works 
equally well at first order,
\begin{equation}
\label{tra1}
\sum_{\kappa} \int d^3 r \, r_\gamma \mathcal{P}_{\alpha, \kappa \beta}({\bf r}) =  \delta_{\alpha \beta} \int d^3r \, r_\gamma \rho^{(0)}({\bf r}).
\end{equation}
On the other hand, we already know from earlier works that the second order doesn't work
if nonlocal potentials are used in the calculation,
\begin{equation}
\label{tra2}
\sum_{\kappa} \int d^3 r \, r_\gamma r_\lambda 
    \mathcal{P}_{\alpha, \kappa \beta}({\bf r}) \neq  \delta_{\alpha \beta} \int d^3r \, r_\gamma r_\lambda \rho^{(0)}({\bf r}).
\end{equation}
This breakdown of translational invariance at the quadrupolar level explains why 
Eq.~(\ref{Eq_mz}) and Eq.~(\ref{Eq_mz2}) disagree in presence of nonlocal potentials.

\section{Structure of molecules used in this work \label{structure}}
\begin{figure}\label{Fig_molecule}
	\includegraphics[width=1.0\linewidth]{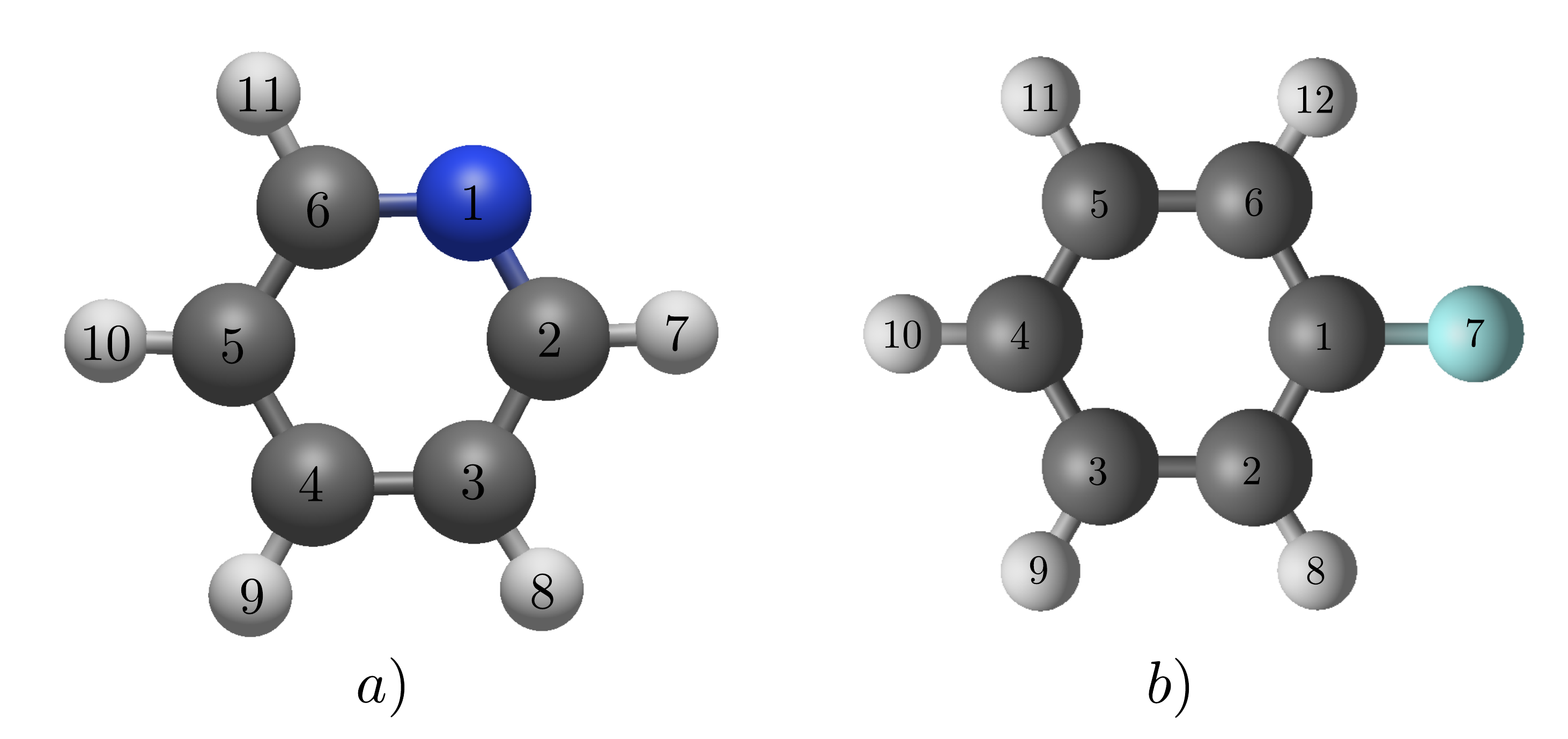}
	\caption{Cartoon illustrating the relaxed geometries used in this work for 
		a) C$_5$H$_5$N (the N atom is in blue) and b) C$_6$H$_5$F (F atom is in light blue).}
	\label{fig:aromatic_compounds}
\end{figure}

Here we show the molecular structures used in this work. For the aromatic compounds, we 
also display a cartoon of the molecules in Figure \ref{Fig_molecule}, labeling each atom with a number. This figure, in combination with Table \ref{tab:molecule_geometry}, enables to construct
the C$_5$H$_5$N and C$_5$H$_5$F molecules. 

In order to calculate the $g$ factor, the rotation axis is taken to be perpendicular to the molecular axis in linear molecules (H$_2$, N$_2$, F$_2$, HF, HNC, and FCCH), perpendicular to the molecular plane for C$_5$H$_5$N, C$_6$H$_5$F and H$_2$O, along one of the bonds in CH$_4$ and perpendicular to the plane formed by the H atoms in NH$_3$; the field is taken to be parallel to the rotation axis.
\begin{table}
	\caption{Molecular geometries for selected simple molecules after relaxation.
		Distances are in bohr.}
	\begin{ruledtabular}
		\label{tab:molecule_geometry}
		\begin{tabular}{c|c}
			&Relaxed geometry \\
			\hline
			H$_2$ & $d=1.446$\\ \hline
			N$_2$ &$d=2.066$\\ \hline
			F$_2$ & $d=2.622$\\ \hline
			HF&$d=1.760$\\ \hline
			\multirow{2}*{HNC}&$d_\text{HN}=1.908$\\
			& $d_\text{NC}=2.202$\\ \hline
			\multirow{3}*{FCCH}& $d_\text{FC}=2.396$\\
			&$d_\text{CC}=2.260$\\
			&$d_\text{CH}=2.022$ \\\hline
			\multirow{2}*{H$_2$O}&$d_\text{HO}=1.835$\\
			&$\angle \text{HOH}=104.8^\circ$\\\hline
			\multirow{2}*{NH$_3$}&$d_\text{NH}=1.930$\\
			&$\angle \text{HNH}=107.3^\circ$\\\hline
			CH$_4$&$d_\text{CH}=2.070$\\\hline	
			\multirow{6}*{C$_5$H$_5$N}&$d_{1,2}=2.509$,
			$d_{2,3}=2.615$\\	
			&$d_{3,4}=2.611$,
			$d_{2,7}=2.071$\\
			&$d_{3,8}=2.063$,	
			$d_{4,9}=2.064$\\
			&$\angle 6,1,2=117.54^\circ$,
			$\angle 1,2,3=123.42^\circ$\\
			&$\angle 2,3,4=118.52^\circ$,
			$\angle 3,4,5=118.58^\circ$\\\hline
			C$_5$H$_5$F&$d_{1,2}=2.603$,
			$d_{2,3}=2.614$\\
			&$d_{3,4}=2.616$,
			$d_{1,7}=2.534$\\
			&$d_{2,8}=2.062$,
			$d_{3,9}=2.063$\\
			&$d_{4,10}=2.062$,
			$\angle 6,1,2=122.43^\circ$\\
			&$\angle 1,2,3=118.40^\circ$,
			$\angle 2,3,4=120.45^\circ$\\
			&$\angle 3,4,5=119.86^\circ$\\
		\end{tabular}
	\end{ruledtabular}
\end{table}

\section{Comparison between DFPT and finite $q$ calculations for cubic SrTiO$_3$ \label{App_comp}}
Here we present a comparison between the  ``DFPT'' implementation described in Sec.~\ref{Sec_implementation_B}, and the ``finite $q$'' implementation described in Sec.~\ref{Sec_implementation_A} for the generalized Lorentz force in cubic SrTiO$_3$. Specifically, we compare the $(D^{\text{di,}z})_{\kappa\alpha,\kappa'\beta}$ elements, see Eq.~(\ref{Eq_Phi_gamma}). All of the independent elements for both methods are presented in Table~\ref{tab:STO_comp}. Note that our labeling convention for the oxygen is, in reduced coordinates: $\text{O}1=(0,1/2,1/2)$, $\text{O}2=(1/2,0,1/2)$, $\text{O}3=(1/2,1/2,0)$.   The additional $(\kappa\alpha,\kappa'\beta)$ can be determined from the following symmetry requirements on the tensor in cubic SrTiO$_3$. For $\kappa$ and $\kappa^\prime$ either (or both) Ti, Sr, or O3, $x\leftrightarrow y$ results in the same magnitude coefficient, with a change of sign. For terms involving O1 and/or O2, exchanging $x\leftrightarrow y$ as well as $\text{O}1\leftrightarrow \text{O}2$ also results in a different sign, but same magnitude coefficient.
Overall, we see quite good agreement, to the second or third decimal places, between the very distinct implementations; this confirms the accuracy of our methodology. 

\begin{table}
	\caption{\label{tab:STO_comp} $(D^{\text{di,}z})_{\kappa\alpha,\kappa'\beta}$ elements [see Eq.~(\ref{Eq_Phi_gamma})] for cubic SrTiO$_3$ calculated with the ``DFPT'' implementation described in Sec.~\ref{Sec_implementation_B}, and the ``finite $q$'' implementation described in Sec.~\ref{Sec_implementation_A}. }
	\begin{ruledtabular}
		\begin{tabular}{crr}
			$(\kappa\alpha,\kappa'\beta)$ & DFPT & Finite $q$ \\\hline
			(Sr $x$, Sr $y$) &$1.5320$ &$1.5317$ \\
			(Sr $x$, Ti $y$)&$2.4483$ &$2.4526$ \\                  
			(Sr $x$, O1 $y$) &$-0.4487$ &$-0.4489$ \\
			(Sr $x$, O2 $y$) &$-2.3829$ &$-2.3874$ \\
			(Sr $x$, O3 $y$) &$-1.1537$ &$-1.1542$ \\
			(Ti $x$, Ti $y$)&$8.4125$ &$8.4037$ \\
			(Ti $x$, O1 $y$)  &$-9.6112$ &$-9.6131$ \\
			(Ti $x$, O2 $y$) &$-2.4061$ &$-2.3937$ \\
			(Ti $x$, O3 $y$) &$3.7315$ &$3.7336$ \\
			(O1 $x$, O2 $y$) &$-6.3272$ &$-6.3416$ \\
			(O2 $x$, O1 $y$) &$3.9400$ &$3.9416$ \\
			(O2 $x$, O2 $y$) &$8.3977$ &$8.3987$ \\
			(O3 $x$, O1 $y$)&$-0.3842$ &$-0.3845$ \\
			(O3 $x$, O2 $y$) &$-2.0026$ &$-2.0041$ \\
			(O3 $x$, O3 $y$)&$0.0642$ &$0.0653$ \\
		\end{tabular}
	\end{ruledtabular}
\end{table}

\end{document}